\documentclass[sigconf]{acmart}
\usepackage{bm}
\usepackage{multirow}
\usepackage{enumitem}
\usepackage[linesnumbered,ruled]{algorithm2e}
\usepackage{makecell}

\newcommand\revise[1]{{\textcolor{black}{#1}}\xspace}
\definecolor{revision}{RGB}{0,0,0}
\newcommand{\revision}[1]{{\color{revision} #1}\xspace}
\newcommand{\revisionstart}{\begin{color}{revision}}
\newcommand{\revisionend}{~\!\!\end{color}}

\def\TR{1} %

\AtBeginDocument{%
  \providecommand\BibTeX{{%
    \normalfont B\kern-0.5em{\scshape i\kern-0.25em b}\kern-0.8em\TeX}}}

\setcopyright{acmcopyright}
\copyrightyear{2023}
\acmYear{2023}
\acmDOI{XXXXXXX.XXXXXXX}

\acmConference[SIGMOD'23]{Proceedings of the 2023ACM SIGMOD International Conference on Management of Data}{June 18--23,2023}{Seattle, WA, USA}
\acmBooktitle{Proceedings of the 2023 ACM SIGMOD International Conference onManagement of Data (SIGMOD'23), June 18--23, 2023, Seattle, WA, USA}
\acmPrice{15.00}
\acmISBN{978-1-4503-XXXX-X/18/06}

\begin{document}
\title{Ground Truth Inference for Weakly Supervised Entity Matching}

\author{Renzhi Wu}
\affiliation{%
  \institution{Georgia Institute of Technology}
}
\email{renzhiwu@gatech.edu}
\author{Alexander Bendeck}
\affiliation{%
  \institution{Georgia Institute of Technology}
}
\email{abendeck3@gatech.edu}
\author{Xu Chu}
\affiliation{%
  \institution{Georgia Institute of Technology}
}
\email{xu.chu@cc.gatech.edu}
\author{Yeye He}
\affiliation{%
  \institution{Microsoft Research}
}
\email{yeyehe@microsoft.com}

\begin{abstract}
Entity matching (EM) refers to the problem of identifying pairs of data records in one or more relational tables that refer to the same entity in the real world. Supervised machine learning (ML) models currently achieve state-of-the-art matching performance; however, they require a large number of labeled examples, which are often expensive or infeasible to obtain. This has inspired us to approach data labeling for EM using \textit{weak supervision}. In particular, we use the labeling function abstraction popularized by Snorkel, where each labeling function (LF) is a user-provided program that can generate many noisy match/non-match labels quickly and cheaply. 
Given a set of user-written LFs, the quality of data labeling depends on a \textit{labeling model} to accurately infer the ground-truth labels.
In this work, we first propose a simple but powerful labeling model for general weak supervision tasks. Then, we tailor the labeling model specifically to the task of entity matching by considering the EM-specific transitivity property.

The general form of our labeling model is simple 
while substantially outperforming the best existing method across ten general weak supervision datasets. %
To tailor the labeling model for EM,   
we formulate an approach to ensure that the final predictions of the labeling model satisfy the transitivity property required in EM, utilizing an exact solution where possible and an ML-based approximation in remaining cases.
On \revise{two} single-table and \revise{nine} two-table real-world EM datasets, we show that our labeling model results in a 9\% higher F1 score on average than the best existing method. We also show that a deep learning EM end model (DeepMatcher) trained on labels generated from our weak supervision approach is comparable to an end model trained using tens of thousands of ground-truth labels, demonstrating that our approach can significantly reduce the labeling efforts required in EM. 
\end{abstract}

\maketitle

\vspace{-1mm}
\section{Introduction}

Entity matching (EM)
refers to the process of determining if a pair of records from two data sources refer to the same real-world entity. EM has many applications, for example, 
in matching product listings for competitive pricing~\cite{competera-product-matching}
and in building knowledge graphs~\cite{DBLP:conf/icde/Dong19}. As a long-standing problem, 
 EM has been extensively studied (e.g., see  surveys~\cite{2007-Springer-EMSurvey-Herzog,2007-TKDE-ElmagarmidDedupSurvey,2012-VLDB-EMGetoorSurvey,2012-Springer-Data-Matching-book,2018-IEEE-Bulletin-Stonebraker.pdf,2018-SIGMOD-DI-ML-Synergy-Tutorial}).

\noindent\textbf{The Need for Weak Supervision in EM.} 
The most important requirement for EM solutions is high matching quality as it directly affects downstream application performance. State-of-the-art EM solutions that report the highest matching quality~\cite{2018-VLDB-DeepER.pdf,2018-SIGMOD-DeepMatcher-Design-Space-Exploration.pdf,2021-VLDB-Ditto,2018-SIGMOD-DI-ML-Synergy-Tutorial} are supervised machine learning (ML) approaches that train a binary classifier to predict the label (match or non-match) for any tuple pair. However, these models require large numbers of labeled tuple pairs, which are often not available or costly to obtain~\cite{2018-SIGMOD-DI-ML-Synergy-Tutorial}. 
The high human cost in data labeling has become a main bottleneck in adopting high-quality EM solutions in practice. %

To meet the needs of data-hungry supervised ML models, ML practitioners have increasingly turned to \textit{weak supervision} methods, in which a larger volume of cheaply generated, but often noisier, labeled examples is used in lieu of hand-labeled examples. Different forms of weak supervision have been investigated, including the use of 
non-expert
crowd workers~\cite{gao2011harnessing}, 
pretrained models~\cite{das2020goggles},
and rules/patterns/heuristics~\cite{shin2015incremental}.

To unify different forms of supervision, the \textit{data programming} paradigm~\cite{2016-NIPS-data-programming.pdf} has been proposed, in which users write \textit{labeling functions} (LFs) to programmatically label training data, rather than manually labeling each example by hand. Each LF is a small user-provided program (e.g., in Python) that leverages noisy signals in the data or domain knowledge to provide a label (or abstain) for an input example. In binary classification tasks such as EM, the output of each LF is +1 (positive class), -1 (negative class), or 0 (abstain). The data programming approach was first implemented in Snorkel~\cite{2018-VLDB-Snorkel-System.pdf}, a weakly supervised data labeling system. Snorkel allows users to write LFs and then uses a generative model to
combine all LFs to produce probabilistic labels. 
The data programming approach has been widely adopted in various ML tasks~\cite{wu2018fonduer, fries2019weakly, lison2020named, wu2022learned, li2021bertifying, fu2020fast, ruhling2021end}.

\begin{figure}[htb!]
\vspace{-4mm}
  \centering
  \includegraphics[width=1\linewidth]{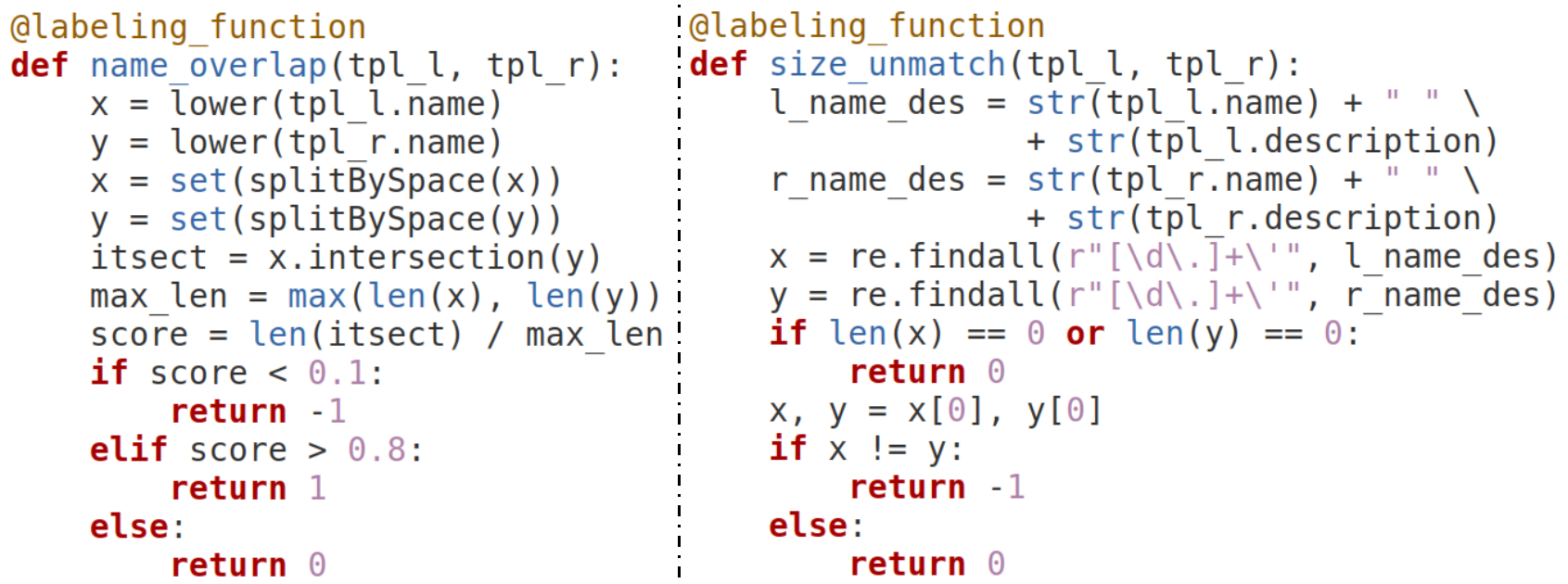}
    \vspace{-6mm}
  \caption{Two user-written LFs for the abt-buy dataset~\cite{erhardws}.}
  \label{fig:example_lfs}
     \vspace{-4mm}
\end{figure}

\noindent\textbf{Example Labeling Functions.} 
The LFs shown in Figure \ref{fig:example_lfs} are two user-written LFs developed using an existing tool~\cite{panda_demo} for matching electronic products~\cite{erhardws}.  
The first LF, "name\_overlap", encodes the intuition that matching products should have similar "name" attributes. Specifically, pairs of products with sufficiently high/low word overlap in their names are predicted as matches (+1)/non-matches (-1).
When the amount of word overlap does not provide conclusive evidence either way, the LF abstains (0). The second LF, "size\_unmatch", uses a regular expression to search the screen size in patterns like "\textcolor{gray}{... Samsung} \textbf{40’} \textcolor{gray}{LCD ...}". %
When the screen sizes of a pair of products are different, the LF predicts the pair to be non-match (-1); otherwise, the LF abstains (0). 

\begin{figure}[htb!]
  \centering
  \includegraphics[width=0.8\linewidth]{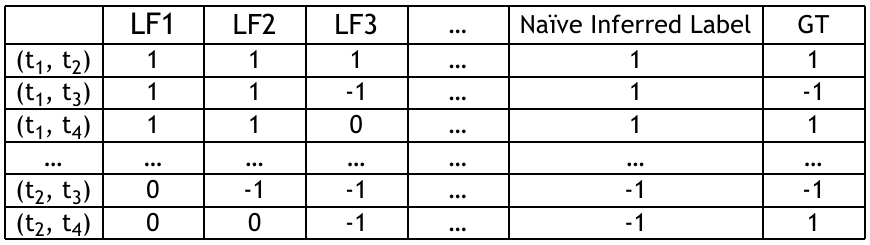}
    \vspace{-4mm}
  \caption{Example of a labeling matrix}%
  \label{fig:labeling_matrix}
     \vspace{-2mm}
\end{figure}

The predictions of multiple LFs can then be used to construct a \textit{labeling matrix} $X$, where each row corresponds to a tuple pair, and each column corresponds to predictions of one LF for all tuple pairs. After constructing the labeling matrix, a \textit{labeling model} is used to consolidate the labeling function predictions into a final label for each tuple pair. Figure \ref{fig:labeling_matrix} shows one instance of a labeling matrix with votes from LFs (LF1, LF2, ...) and inferred labels from a naive labeling model (+1 for match, -1 for non-match) as well as the unknown ground truth (GT) labels for each tuple pair.

Given a set of user-written LFs, the quality of data labeling depends on the \textit{labeling model} that infers the ground-truth labels from the LFs.
In this paper, we first present an embarrassingly simple yet highly effective labeling model (SIMPLE) for general weak supervision tasks. 
We then specialize the labeling model (SIMPLE-EM) for the task of EM by considering the transitivity property. 
\begin{sloppypar}
\noindent\textit{\underline{(1) Labeling Model for General Weak Supervision Tasks:}} Existing methods are mostly generative models, i.e., they model the process of generating the predictions of each LF from the hidden ground-truth labels~\cite{ratner2016data,dawid1979maximum, fu2020fast, 2019-ICML-Exploiting-Worker-Correlation-for-Label-Aggregation.pdf}.
Under an assumed generative process, one could measure the likelihood of observing all LFs (i.e., the labeling matrix) and an estimation of the ground-truth labels can be obtained by maximizing the likelihood~\cite{ratner2016data,dawid1979maximum, fu2020fast,2019-ICML-Exploiting-Worker-Correlation-for-Label-Aggregation.pdf}. A fundamental limitation of these approaches is the need to assume that the generative process takes certain simplified forms for ease of modeling or to make the model mathematically solvable. Such assumptions include the Markov 
assumption~\cite{dawid1979maximum, ratner2016data, fu2020fast} and the mixture of independent sub-types assumption~\cite{2019-ICML-Exploiting-Worker-Correlation-for-Label-Aggregation.pdf}. 
Evidently, these assumptions can often break in practice. Recent benchmark studies~\cite{zhang2021wrench} suggest that the best existing labeling model for general weak supervision tasks is only 1.5\% better than majority vote, %
 and the second best method is already worse than majority vote.
In addition, even with these various assumptions to simplify the generative process, the existing approaches are typically quite complicated and difficult to implement. 
\end{sloppypar}

We take a fundamentally different approach that is a significant departure from the conventional wisdom that generative models should be used in labeling models.
The core intuition of our method is that \textit{labeling models are functions}. 
Conceptually, every labeling model takes in the labeling matrix $X$ and predicts the label vector $\bm{\hat{y}}$.
For example, $X$ could be the sub-matrix in Figure~\ref{fig:labeling_matrix} with columns of all LFs and $\bm{\hat{y}}$ could be the naive inferred label column in Figure~\ref{fig:labeling_matrix}.
Designing a labeling model is to design a function $G$ parameterized by $\Theta$, and $\bm{\hat{y}}=G(X, \Theta)$ would be the predicted labels of the model. 
\revise{
The prediction of the whole matrix $X$ can be made by independently predicting each individual row/data point; i.e., $\hat{y_i}=g(X[i,:], \theta)$ where $\theta$ is the parameter of $g$. Different designs of existing labeling models can be seen as the function $g$ taking different forms.}

As discussed, existing labeling models typically build complicated models (e.g., probabilistic graphical models~\cite{demartini2012zencrowd, karger2011iterative, liu2012variational} and matrix completion models~\cite{ratner2019training, ibrahim2021crowdsourcing}) by making various assumptions to handcraft $g$.
In this paper, we ask a fundamental question that is contrary to the conventional wisdom: \textit{Can we use a generic classifier (e.g., random forest) as $g$ to be the labeling model and avoid all the complicated designs in the existing approaches?} We provide an affirmative answer to this question by presenting an embarrassingly simple method (in Section~\ref{sec:proposed_truth_inference}) that achieves better performance than the existing approaches across ten datasets.%

\begin{sloppypar}
\noindent\textit{\underline{(2) EM-specific Labeling Model with Transitivity:}} The \textit{transitivity property of EM} states that for any three tuples $t_i$, $t_j$, and $t_k$, if $t_i$ matches $t_j$ and also matches $t_k$, then $t_j$ has to match $t_k$. 
Concretely, consider the scenario in Figure \ref{fig:labeling_matrix} -- a naive model would likely predict the tuple pair $(t_2, t_4)$ as non-match,
 despite strong evidence that both $(t_1, t_2)$ and $(t_1, t_4)$ should be matches, which results in inconsistent label assignments that violate transitivity. Our goal is to explicitly model transitivity using constraints, so that matching decisions can be made in a holistic manner. %
\end{sloppypar}
A naive approach to enforce transitivity is to apply it as a post-processing step, which is sub-optimal.
Intuitively, the additional signals from the transitivity constraints should be made an integral part of the labeling model to improve its accuracy.
We thus attempt to incorporate transitivity directly in the labeling model. However, this makes the labeling model complex and difficult to solve.

To tackle this challenge, we first consider a simplified yet common scenario in two-table EM, where at least one table is ``duplicate-free'' (this \revise{has been} shown to hold for many real-world datasets~\cite{AutoFJ} \revise{which we further verify in Section \ref{ssec:study_trans}}).
In this scenario, we can derive an exact solution incorporating transitivity in the labeling model. 
In the more general setting where
neither table is duplicate-free or in single-table EM, we propose an ML-based approach (in Section \ref{tup-trans-section}) that produces an approximate solution with transitivity incorporated into the labeling model. 
We highlight that 
our ML model can be trained offline once and then used for any new EM datasets without needing any form of update.%

\noindent\textbf{Contributions.} In this work, 
we make the following contributions:
\begin{itemize}[leftmargin=0mm]
\vspace{-1mm}
    \item[] \textit{(1) A generic view of labeling models.} We present a generic view of labeling models as functions (or classifiers). Different existing methods are simply various instantiations of this generic view. 
    \item[] \textit{(2) A simple and powerful labeling model.}
    Based on the generic view, 
    we study a fundamental question: Is it possible to use a generic classifier 
     as the labeling model? We provide an affirmative answer by proposing a simple yet powerful labeling model based on random forest. Our method achieves better results than existing approaches across ten general weak supervision datasets.%
     \item[] \textit{(3) Exact solution for transitivity. }
     We adapt our proposed labeling model to the task of two-table entity matching by considering the transitivity property. 
     We present an exact solution to transitivity when at least one table is duplicate-free for two-table EM. %
      \item[] \textit{(4) ML-based solution for transitivity.}
      We adapt our proposed labeling model to the task of single-table entity matching by considering the transitivity property. We propose an ML-based approach based on the observation that, in an optimization step of the labeling model, the optimal matching probabilities under the transitivity constraint depend only on the optimal probabilities without considering the transitivity constraint. We further design a novel architecture for the ML model by considering task-specific properties. Our ML model trained once offline can be applied to any new EM datasets without needing any adaptation.

\end{itemize}

\vspace{-2mm}
\section{Problem Definition}
Given $m$ labeling functions (LFs) and $n$ data points 
to be labeled, the \textit{labeling matrix} $X \in \{+1,-1,0\}^{n \times m}$ is obtained by applying all LFs to all data points. Since LFs are noisy and may have dependencies (leading to overlaps and conﬂicts in $X$), our goal is to design and implement a \textit{labeling model} to optimally combine the LF labels in $X$ to obtain accurate final labeling results. 

For the task of entity matching, each data point is a tuple pair $(t_i, t_j)$ and the predicted label is denoted as $\hat{y}^{(i,j)}\in\{-1, +1\}$. 
The final predictions must satisfy the transitivity constraint that if $\hat{y}^{(i,j)}$ and $\hat{y}^{(i,k)}$ are both equal to +1 (denoting matches), then $\hat{y}^{(j,k)}$ must also be +1. \revise{This form of the transitivity constraint applies to single-table EM; for two-table EM, the matching probabilities of tuple pairs from the same table are typically not available, so in practice transitivity relies on one or both tables being duplicate-free.}

\vspace{-2mm}
\section{Proposed Labeling Model}
\label{sec:proposed_truth_inference}
In this section, we introduce the general form of our proposed labeling model for general weak supervision tasks. 

\vspace{-2mm}
\subsection{A Generic View of Labeling Models}
\label{ssec:generic_view}
Every labeling model (also referred to as truth inference method) takes in the labeling matrix $X$ and predicts the label vector $\bm{\hat{y}}$. At a high level, we can think of a truth inference algorithm as a function $G$ parameterized by $\Theta$ such that $\bm{\hat{y}}=G(X, \Theta)$. In most tasks, the data points are independent, so the predictions of their labels are also made independently. Therefore, the function $G$ that makes predictions for the whole labeling matrix can be expressed with a function $g$ that makes a prediction for each individual data point. When predicting the $i^{th}$ data point, the function $g$ takes in the $i^{th}$ weak label vector (i.e., the $i^{th}$ row in matrix $X$) $x_i=X[i,:]$ as features and predicts the probability of being in the positive class $\gamma_i$, i.e. $g(x_i, \theta)$ 
where $\theta$ is the parameter of $g$. The label $\hat{y}_i$ is obtained as $\hat{y}_i=1$ if $g(x_i, \theta) \geq 0.5$ and $\hat{y}_i=0$ otherwise.
In this formulation, $g$ is essentially a classifier that takes in a feature vector $x_i$ and predicts a soft label. 
In principle, $g$ could be any classifier, and the challenge is how to learn its parameter $\theta$ without labeled data. 

\noindent\textbf{Expectation-Maximization Algorithm.} Most labeling models (or truth inference methods) adopt the Expectation-Maximization algorithm (or extensions of it) to learn the model parameter $\theta$. The objective function 
is the negative log data likelihood function:
\begin{equation}
\label{eq:EM_obj}
L(\theta, X, \bm{\gamma}) = -\Sigma_{i=1}^N \big( \gamma_i\log(g(x_i, \theta))+(1-\gamma_i)\log(1-g(x_i, \theta)) \big)
\end{equation}
where $\bm{\gamma}=\{\gamma_1, \dots\}$ is the hidden ground-truth label. Note we use $\bm{\gamma}$ to denote the soft labels (the matching probabilities) and $\bm{\hat{y}}$ to denote the hard labels. 
Minimizing the negative log data likelihood function translates to maximizing the likelihood of observing the labeling matrix $X$. 
Since both the parameter $\theta$ and label $\bm{\gamma}$ are unknown, the Expectation-Maximization algorithm iteratively estimates  $\theta$ by minimizing the objective function with $\bm{\gamma}$ fixed as the current estimation and then computes $\bm{\gamma}$ using the current estimated $\theta$.
Specifically, 
the learning workflow of Expectation-Maximization algorithm is as follows:
\vspace{-1mm}
\begin{enumerate}[label=(\arabic*),leftmargin=*]
\item Obtain an initial estimation of the hidden ground-truth label $\bm{\gamma}$ by majority vote, as this works well in practice~\cite{zhang2021wrench}.
\item \textit{M-step}: Estimate the model parameter $\theta$ by minimizing Equation~\ref{eq:EM_obj} with respect to $\theta$ while keeping the labels $\bm{\gamma}$ fixed as the current estimation. This uses the current estimated labels to learn the model parameters.
\item \textit{E-step}: Update the estimated labels $\bm{\gamma}$ as the predicted matching probabilities $\gamma_i = g(x_i, \theta), \ \forall i$ using the model parameter $\theta$ obtained in the M-step.
\item Repeat steps (2) and (3) until convergence. 
\end{enumerate}
\vspace{-1mm}

It can be easily shown (by taking derivative and setting it equal to zero) that in the M-step, the global minimum of Equation~\ref{eq:EM_obj} 
with respect to $\theta$ is achieved when $g(x_i, \theta)=\gamma_i$. 
This means the objective function measures the difference/loss between the model prediction and the current soft label, and thus one can replace the objective function in M-step to be any loss function $D(\gamma_i, g(x_i, \theta))$ which also has a minimum at $g(x_i, \theta)=\gamma_i$ to achieve similar results. In other words, in the M-step, we can update the model parameter $\theta$ by minimizing $ \frac{1}{N}\sum_{i=1}^N D(\gamma_i, g(x_i, \theta))$ where $N$ is the number of examples. Since some classifiers require specific forms for their loss functions (e.g., SVM and random forest), the possibility to substitute the objective function with any loss function while achieving similar results in the M-step enables us to instantiate $g$ to be any classifier. 

\noindent\textbf{Instantiation as Different Labeling Models.}
By choosing different forms of $g(x_i, \theta)$, the Expectation-Maximization algorithm is instantiated as different existing labeling models.
 The existing methods design $g(x_i, \theta)$ by making different assumptions.

One common assumption is that the noisy labels of each LF are only dependent on the ground-truth labels, so the noise can be modeled by the joint probability table (the confusion matrix) of the LF and the hidden ground-truth. Based on this, a family of truth inference methods~\cite{dawid1979maximum, raykar2010learning, venanzi2014community} has been developed. These models can be obtained by instantiating $g(x_i, \theta)$ as the probability distribution of the class labels derived based on the confusion matrix.

Another common assumption is the Markov assumption.
This is the core assumption in probabilistic graphical models (PGM)~\cite{koller2009probabilistic}. With this assumption, the truth inference problem can be formulated as a PGM with hidden variables. A rich family of truth inference methods has been developed using PGMs. Such models can be obtained if we represent $g(x_i, \theta)$ with a PGM. Different methods in this family differentiate from each other by adopting different priors~\cite{demartini2012zencrowd}, or by replacing the Expectation-Maximization algorithm with its extension (Variational Bayes~\cite{Variational_bayesian})~\cite{karger2011iterative, liu2012variational}.

\vspace{-3mm}
\subsection{The SIMPLE Algorithm.}
In principle, $g$ can be any classifier. Our intuition for choosing a proper $g$ is that the capacity (also referred to as complexity or size of the hypothesis space in the literature) of $g$ should not be too large, otherwise $g$ will learn a trivial solution (the prediction of majority vote) in the M-step of the first iteration. On the other hand, the capacity of $g$ should not be too small as we need $g$ to capture the interaction and dependency of different LFs (features).

In fact, different existing methods can be viewed as restricting the capacity of $g$ in different ways through different intuitions or assumptions. Specifically, by designing $g$ based on the assumptions (e.g., conditional independence), these methods implicitly restrict the space of $g$ to only include functions that observe the assumptions, 
so the model avoids learning the trivial solution (the initialized labels, e.g., by majority vote) and can generalize well.%

From this view, one natural question is the following: 
Is it really necessary to handcraft complicated models with various assumptions to limit capacity? Or can we directly use a generic classifier as $g$ and explicitly limit its capacity to achieve similar results? The answer to this latter question turns out to be yes. 
We show that we can actually use a generic classifier (random forest) as $g$ and achieve better performance by explicitly restricting its capacity. 

To use a generic classifier as $g$, a straightforward choice is to use a simple linear classifier (i.e., logistic regression). This reduces to the weighted majority vote method. In other words, logistic regression assigns a weight for each LF and then combines all LFs using the weights to get the final label. The limitation of logistic regression is that it is not able to capture more complex interactions or dependencies between different features (LFs). 

To choose a proper classifier, our intuition is that we want a classifier (1) that is able to express the interaction of different LFs, and (2) for which we can easily restrict its capacity. Tree-based approaches naturally model the interaction of different features (LFs). For example, in a decision tree, the model makes decisions based on different features at different levels of the tree, so the tree as a whole naturally considers the interactions of different features (LFs).
Tree-based approaches also work very well on structured data in practice. In fact, a recent survey reveals that tree-based methods are still the most common winning solutions in ML competitions on structured data~\cite{Olaleye2022Mar}. Therefore, we choose the random forest classifier, a classic tree-based method, for our task. Note that we can easily restrict the capacity of a random forest classifier by setting its hyper-parameters. 
The first parameter is the maximum tree depth $d_{\text{max}}$. 
With a smaller maximum depth, the random forest classifier has smaller capacity.
Another parameter that controls the overall capacity/complexity of the trees is the complexity parameter: ccp\_alpha. Following the common practice, we select both parameters $d_{\text{max}}$ and ccp\_alpha using cross validation.
Note cross validation is done with the current estimated labels at each M-step where we train the classifier and no ground-truth labels are used.

Restricting the capacity of a classifier is also known as regularization. In our method, regularization is done as in typical ML tasks -- in a way that is explicit and data-driven through cross validation. In contrast, in existing truth inference methods, regularization is done implicitly through manually restricting the hypothesis space (e.g., the form of function $g$) based on various assumptions or heuristics.

With a random forest classifier as $g$, the M-step in the Expectation-Maximization algorithm simply becomes training the classifier with the current estimated labels (we use the hard labels $\hat{\bm{y}}$ obtained by binarizing the soft labels $\bm{\gamma}$ because common implementations of random forest only support training with hard labels), and the E-step simply becomes performing prediction on the training set to obtain an updated version of the soft labels $\bm{\gamma}$.

\noindent\textbf{Class Imbalance.} 
Many real world datasets (especially EM datasets) have imbalanced classes. 
Traditionally, to handle the data imbalance problem, one would need to carefully design the labeling model (truth inference method) such as by introducing priors~\cite{2018-VLDB-Snorkel-System.pdf, fu2020fast, 2019-ICML-Exploiting-Worker-Correlation-for-Label-Aggregation.pdf}. In our method, handling the problem of class imbalance is the same as handling class imbalance in a typical ML setting, and we are able to directly adopt state-of-the-art techniques to address it. 

We handle the class imbalance problem when training the classifier at the M-step. Specifically, we augment the data points of the minority class to match the size of the majority class with SMOTE~\cite{chawla2002smote}, a simple technique for class imbalance that works very well in practice.  SMOTE~\cite{chawla2002smote} works by creating synthetic examples by interpolating existing data points. For example, if we have two data points in the positive class $(x_1, 1), (x_2, 1)$, SMOTE might create one synthetic data point by interpolating the two data points, e.g., $(\frac{x_1+x_2}{2}, 1)$.
We train the model with the augmented minority class and the original majority class at each M-step.  After training, in the E-step, we perform prediction on the original set of data points to get a new version of the labels $\bm{\gamma}$. Since we do not know whether a given dataset has the class imbalance problem or not, we always apply SMOTE at each M-step for all datasets.
The pseudo-code of the entire SIMPLE algorithm is shown in Algorithm~\ref{alg:simple}. 
We open-source our implementation at~\cite{SIMPLE_code}.

\setlength{\textfloatsep}{0pt}
\begin{algorithm}
\caption{SIMPLE}
\label{alg:simple}
\SetAlgoLined
\KwIn{Labeling matrix $X$}
\KwOut{Estimated soft labels $\bm{\gamma}$}
$\bm{\gamma} \gets$ majority vote on $X$\\
\While{Not Converged}{ \label{algo_line:termination}
\textbf{M Step}\\  \label{algo_line:m_step_start}
\Indp Obtain hard labels $\hat{\bm{y}}$ by binarilize the soft labels $\bm{\gamma}$.\\
Make the classes balanced: $X',\hat{\bm{y}}' = \text{SMOTE}(X, \hat{\bm{y}})$\\
Select random forest parameters $d_{\text{max}}$ and ccp\_alpha with cross validation on data $(X',\hat{\bm{y}}')$ \\
RandomForestClassifier.fit$(X',\hat{\bm{y}}')$\\
\label{algo_line:M_end}
\Indm
\textbf{E Step}\\
\Indp $\bm{\gamma} \gets$  RandomForestClassifier.predict\_proba$(X)$\\ \label{algo_line:e_step}
\Indm
}
\Return $\bm{\gamma}$
\end{algorithm}

\noindent\revise{\textbf{Computational complexity.} The complexity of each iteration is dominated by training the random forest classifier, which has a time complexity of $O(N\log(N))$ where $N$ is the number of tuple pairs in the candidate set~\cite{louppe2014understanding}. Let $M_I$ denote the number of iterations. The overall time complexity is $O(M_IN\log(N))$. In our experiments, we observe that 10 iterations is enough for all datasets. The space complexity is $O(N)$.
}

\noindent\textbf{Discussion.} Our method has a connection to the pseudo-labeling method in semi-supervised learning~\cite{lee2013pseudo, arazo2020pseudo}. In pseudo-labeling, first a model is trained on a labeled training set and used to make predictions on unlabeled data; the data with confident predicted labels are then added to the training set. Next, the model is trained on the new training set, and then this process continues iteratively. Both pseudo-labeling and our method use predicted data as labeled data for the next iteration. The differences between pseudo-labeling and our method are as the following: First, pseudo-labeling is in a semi-supervised setting while our method is in a unsupervised setting. We highlight that truth inference might be the only (or at least one of the few) unsupervised task(s) where one could apply this type of iterative approach as one can get a reasonable initial estimation (e.g., by majority vote in truth inference) to start with. This is due to the implicit assumption in weak supervision that each user-provided LF is better than random guessing. For general unsupervised ML tasks, there is no straightforward way to get a good initial estimation.
Second, the focus of pseudo-labeling is to train a model while our focus is to obtain the labels for all data points; consequently, pseudo-labeling only adds confident label predictions to the training set while we include all predictions. 

\vspace{-1mm}
\section{Incorporating Transitivity} \label{tup-trans-section}
In this section, we tailor our proposed labeling model to the task of EM by incorporating the transitivity property. The tailored method is denoted as SIMPLE-EM. 
The transitivity property for EM states that if tuple $t_i$ matches $t_j$ and $t_i$ matches $t_k$, we must conclude that $t_j$ also matches $t_k$. 
We follow prior work~\cite{ZeroER} to model transitivity as an inequality constraint, and the E-step of the EM algorithm can then be formulated as a constrained optimization problem. To make this section self-contained, we briefly review the formulation from prior work~\cite{ZeroER} in Section~\ref{ssec:constrained_formulation}. Then, we introduce our solution in Section~\ref{ssec:trans_two_table} and Section~\ref{ssec:trans_single_table}. 

\subsection{Constrained Optimization Formulation for Transitivity.}
\label{ssec:constrained_formulation}
\noindent\textbf{Transitivity as Constraint on Matching Probabilities.}
We follow prior work~\cite{ZeroER} to model transitivity as an inequality constraint defined on matching probabilities. Specifically, for any three tuples $t_i$, $t_j$, and $t_k$, the transitivity constraint can be expressed as the following inequality:
\vspace{-2mm}
\begin{equation}
\label{eq:transitivity_as_inequality}
\begin{small}
\gamma^{(i,j)} \times \gamma^{(i,k)} \leq \gamma^{(j,k)}
\end{small}
\vspace{-2mm}
\end{equation}
where we use the superscript ${(i,j)}$ (e.g., $\gamma^{(i,j)}$) to index the tuple pair $(t_i,t_j)$ and $\gamma^{(i,j)}$ denotes the matching probability of $(t_i,t_j)$. To see how Equation~\ref{eq:transitivity_as_inequality} secures the transitivity constraint, consider an example where $\gamma^{(i,j)}=0.6$ and $\gamma^{(i,k)}=0.5$: in $60\%$ of the cases $(t_i, t_j)$ is a match and in $50\%$ of the cases $(t_i, t_k)$ is a match. Transitivity applies only when both $(t_i, t_j)$ and $(t_i, t_k)$ are matches which is in $60\%\times 50\%=30\%$ of the cases, and in these cases $(t_j, t_k)$ is a match due to transitivity. Therefore, $(t_j, t_k)$ has at least a $30\%$ chance of being a match, as captured by Equation~\ref{eq:transitivity_as_inequality}. 

The set of all transitivity constraints defines a \textit{feasibility set} for $\boldsymbol{\gamma}$: $Q = \{\boldsymbol{\gamma}|\gamma^{(i,j)}\gamma^{(i,k)} \leq \gamma^{(j,k)}   \forall i,j,k \}$.
Intuitively, incorporating transitivity reduces to ensuring that the labeling model's prediction $\bm{\gamma}$ is in the feasibility set $Q$. 

\noindent\textbf{Incorporating Constraints in EM.}
On the surface, it seems very difficult to incorporate the transitivity constraints $Q$ into the labeling model -- namely, to find the best model parameter $\theta$ that minimizes $L(\theta, X, \bm{\gamma})$, while ensuring that the probabilities $\bm{\gamma}$ that are directly computed based on $\theta$ in the E-step satisfy $Q$. Following prior work~\cite{ZeroER}, this is solved by using the free energy view of the expectation-maximization algorithm~\cite{neal1998view}. In this view, the objective function becomes the negative free energy function $F(\theta, X, \boldsymbol{\gamma})$: 
\begin{small}
\vspace{-2mm}
\begin{equation}
\label{eq:free_energy}
        F(\theta, X, \boldsymbol{\gamma}) = \sum_{(i,j)} -\gamma^{(i,j)}\log\frac{g(x^{(i,j)}, \theta)}{\gamma^{(i,j)}}
        -(1-\gamma^{(i,j)})\log\frac{1-g(x^{(i,j)}, \theta)}{1-\gamma^{(i,j)}}
\end{equation}
\end{small}
The M-step stays the same as it was before, while the E-step also becomes an optimization process that allows us to incorporate the transitivity constraints. Specifically, the E-step becomes $\bm{\gamma}^* = \arg \min_{\boldsymbol{\gamma}} F(\theta, X, \boldsymbol{\gamma})$ which can be shown to be equivalent to the original E-step where $\bm{\gamma}^*$ is obtained by direct computation (i.e., $\gamma^{*(i,j)} = g(x^{(i,j)}, \theta)$)~\cite{neal1998view}. With this formulation, the transitivity constraint can be incorporated in E-step by constraining $\bm{\gamma}\in Q$:
\begin{equation}
\label{eq:constrained_opt}
\small
\boldsymbol{\gamma}^{**} = \arg \text{min}_{\boldsymbol{\gamma} \in Q} F(\theta, X, \boldsymbol{\gamma})
\end{equation}
Intuitively, $\boldsymbol{\gamma}^{**}$ would be the best matching probabilities that satisfy the transitivity constraint while minimizing the objective function. 
However, the above constrained optimization problem is difficult to solve since the constraint set $Q$ is actually non-convex (the Hessian matrix of a constraint is indefinite)~\cite{ZeroER}. Existing work uses a projection-based heuristic~\cite{ZeroER} to address this, which is not robust across datasets as we show in experiments. We will propose a more principled and efficient solution.%

\noindent\textbf{Relationship Between Constrained and Unconstrained Solution.} The observation of our solution is that the constrained solution $\boldsymbol{\gamma}^{**}$ is only dependent on the unconstrained solution $\boldsymbol{\gamma}^{*}$; in other words, there exists a function $h$ such that $\boldsymbol{\gamma}^{**}=h(\boldsymbol{\gamma}^{*})$. To see this,  $\boldsymbol{\gamma}^{*}$ is obtained as $\gamma^{*(i,j)} = g(x^{(i,j)}, \theta)$ in the E-step and we can replace $g(x^{(i,j)}, \theta)$ with $\gamma^{*(i,j)}$ in the objective function in Equation~\ref{eq:free_energy}. In this way, the only two variables in the objective function are $\gamma^{*(i,j)}$ (which is known) and $\gamma^{(i,j)}$ (which is to be solved and the solution is denoted as $\gamma^{**(i,j)}$).
Since $\boldsymbol{\gamma}^{*}$ and $\boldsymbol{\gamma}$ are the only two variables in the objective function and the constraint of Equation~\ref{eq:constrained_opt},
the optimal solution of $\boldsymbol{\gamma}$ (i.e., $\boldsymbol{\gamma}^{**}$) is only dependent on $\boldsymbol{\gamma}^{*}$.

\vspace{-3mm}
\subsection{Transitivity for Two-Table EM}
\label{ssec:trans_two_table}
For two-table EM, it has been found that in most real-world datasets at least one of the two tables is duplicate-free~\cite{AutoFJ}. Under the scenario that one or both tables are duplicate-free, we can derive the exact $\boldsymbol{\gamma}^{**}$ from $\boldsymbol{\gamma}^*$. (Again, note that $\bm{\gamma}^*$ can be easily obtained by direct computation in the E-step.)

\noindent\textbf{One Table Is Duplicate-free.}
Assume without loss of generality that the left table is known to be duplicate-free. For any tuple pair $(t_{l_i}, t_{l_j})$ from the left table, the duplicate-free information is incorporated into the model formulation by setting $\gamma^{**(l_i, l_j)}=\gamma^{*(l_i, l_j)}=0$. We can show that in the constrained solution $\boldsymbol{\gamma}^{**}$, for any tuple $t_{r_k}$ from the right table, there exists only one tuple from the left table that has non-zero matching probability with $t_{r_k}$. To see this by contradiction, let $t_{l_i}$ and $t_{l_j}$ denote two tuples from the left table that have non-zero matching probability with $t_{r_k}$. By the transitivity constraint, $\gamma^{**(r_k, l_i)}\gamma^{**(r_k, l_j)} \leq \gamma^{**(l_i, l_j)}=0$, which means at least one of $\gamma^{**(r_k, l_i)}$ and $\gamma^{**(r_k, l_j)}$ must be zero, contradicting the fact that $t_{l_i}$ and $t_{l_j}$ both have non-zero matching probability with $t_{r_k}$.

Assume that we have the optimal unconstrained matching probabilities $\gamma_k^*=\{\gamma^{*(r_k, l_1)}, \gamma^{*(r_k, l_2)}, \dots\}$ between $t_{r_k}$ and the left table tuples. Since the constrained solution can only have one of these probabilities be nonzero, the optimal solution under the transitivity constraint $\gamma_k^{**}$ is obtained by setting $|\gamma_k^*|-1$ values to be zero in $\gamma_k^*$. We would like to keep as nonzero the variable with the maximum increase in the objective function when it is set to zero, so that the overall objective is minimized. The objective function increase of a single probability $\gamma^{*(r_k, l_i)}$ is obtained as $\Delta F(\gamma^{*(r_k, l_i)})=\log(1/(1-\gamma^{*(r_k, l_i)}))$ which is monotonic to $\gamma^{*(r_k, l_i)}$. As a result, the algorithm of obtaining $\boldsymbol{\gamma}^{**}$ from $\boldsymbol{\gamma}^*$ is as follows: For every tuple in the right table, find the left tuple with the maximum probability of matching to it, and set the matching probabilities of all other left tuples to be zero. 
\revise{The time complexity of this step is $O(N)$ and the time complexity of the overall SIMPLE-EM algorithm is still $O(M_IN\log(N))$. Note this method is the same as the one used in SiGMa~\cite{lacoste2013sigma} and LINDA~\cite{bohm2012linda}. In prior work~\cite{lacoste2013sigma, bohm2012linda} the method was used as a greedy solution for the general setting, while we point out that in our formulation it is the optimal method in the case when one table is duplicate-free. When two tables are both duplicate-free, this method is sub-optimal/greedy and we propose a different solution in the following paragraph.}

\noindent\textbf{Two Tables Are Duplicate-free.} When both tables ($L$ table and $R$ table) are known to be duplicate free, we can follow the same reasoning as in the one-table duplicate-free case and extend it to be bi-directional. Thus, every tuple in the left table can have non-zero matching probability to only one tuple in the right table, and every tuple in the right table can have non-zero matching probability to only one tuple in the left table. Therefore, for all $|L|\times |R|$ possible left tuple and right tuple pairs, we want to keep $\min(|L|,|R|)$ pairs and set the matching probabilities of all other pairs to be zero. In particular, we would like to keep the $\min(|L|,|R|)$ pairs with the minimum  objective function values while satisfying the condition that every left (or right) tuple can have non-zero matching probability to at most one tuple in the right (or left) table. \revise{This is essentially the \textit{assignment problem}, and there is an existing efficient algorithm to solve it -- the LAPJV algorithm ~\cite{jonker1987shortest, cui2016solving} with a time complexity of $O(N\min(N_l, N_r))$. 
Note that the time complexity listed in the original paper~\cite{jonker1987shortest} is the dense version which corresponds to the setting without blocking. The complexity $O(N\min(N_l, N_r))$ is the sparse case with blocking and can be derived following~\cite{cui2016solving}. 
Since we only care about matches, we only need to consider the pairs with a matching probability greater than 0.5. In this case, the time complexity can be further optimized to be  $O(N_{M}\min(N_{l,M}, N_{r,M}))$, where $N_{M}$ is the number of predicted matches (typically orders of magnitude smaller than $N$) and $N_{l,M}$ and $N_{r,M}$ are the number of left and right tuples involved in the predicted matches. 
In our experiments, we do not adopt this optimization as we empirically observed that the algorithm finishes in a reasonable time without the optimization.
The time complexity of the SIMPLE-EM algorithm in this case is $O(M_IN(\log(N)+\min(N_l, N_r))$.
The space complexity is $O(N)$.
We use an existing efficient implementation of the LAPJV algorithm in the scipy package
~\cite{scipy_assign}.}

\revise{
Since the LAPJV algorithm is non-trivial and technically dense, here we only briefly introduce the high-level ideas of the algorithm.  
The LAPJV algorithm first reformulates the assignment problem as a minimum cost flow problem and then solves it by finding the shortest path on an auxiliary graph~\cite{jonker1987shortest}. It further employs several techniques (e.g., column reduction, reduction transfer, and augmenting row reduction) to quickly filter out unlikely paths~\cite{jonker1987shortest}. For more details, one can refer to the original paper~\cite{jonker1987shortest}.%
}

\noindent\textbf{Duplicate-free Detection with Weak Supervision.}
We use the exact solutions outlined above when it is known one or two tables is duplicate-free.
However, in some cases, the information of whether one table is duplicate-free is unknown.
Therefore, we propose a method that leverages the results of the labeling model (without considering transitivity) that labels left-right (LR) tuple pairs with LFs to detect whether either table is duplicate free. The intuition of our detection method is that, if the left table is not duplicate-free, one right-table tuple might appear in multiple LR matching tuple pairs. For example,  $(t_{l_1}, t_{r_1})$ and $(t_{l_2}, t_{r_1})$ are both matches; due to the fact the left table contains duplicates $t_{l_1}$ and $t_{l_2}$, the right tuple $t_{r_1}$ appeared twice in the matching pairs. On the other hand, when the left table is duplicate-free, one right-table tuple can only appear once in the LR matching tuple pairs. This means the distribution of the right-table tuples in the LR matching tuple pairs is different in the two cases.
The predicted LR matching tuples pairs from the labeling model might be noisy, but still provide some information that we can use to detect which of the two cases the distribution of the right-table tuples falls in by using a hypothesis test procedure.
\if\TR 1
We describe and experimentally evaluate the method in Appendix~\ref{app:dup_free} due to space limit. 
\else
We provide detailed description and experimental evaluation of the method in a technical
report~\cite{url:technical_report} due to space considerations.
\fi
We highlight that the proposed duplicate-free detection method does not require LFs for left-left (LL) or right-right (RR) tuple pairs. Since the user has already written LFs for LR pairs, the method requires no additional effort from the user.

\vspace{-1mm}
\subsection{Transitivity for Single-Table EM}
\label{ssec:trans_single_table}
For single-table EM, we are not able to leverage the duplicate-free information to derive the exact constrained solution $\boldsymbol{\gamma}^{**}$ from the unconstrained solution $\boldsymbol{\gamma}^*$. Since we showed that $\boldsymbol{\gamma}^{**}=h(\boldsymbol{\gamma}^*)$ at the end of Section~\ref{ssec:constrained_formulation}, we propose to train a model offline to approximate $h$. Specifically, we randomly generate many instances of $\boldsymbol{\gamma}^*$, and employ expensive numerical solvers to obtain the corresponding $\boldsymbol{\gamma}^{**}$. In this way, we obtain many pairs of $(\boldsymbol{\gamma}^*, \boldsymbol{\gamma}^{**})$, which are used as training data to train an ML model to approximate $h$. 

\noindent\revise{
\textbf{Trained model will be dataset-agnostic.}
Before we go into the details of how we train the model, we emphasize that the trained model will also work on unseen datasets that may differ from the generated training set. To see this, consider that the form of the function $h$ is dataset-independent as no dataset-specific information is involved in our derivation at the end of Section~\ref{ssec:constrained_formulation}. If we can derive the analytical form of $h$, the analytical form can surely be used for any dataset. 
However, obtaining the analytical form is difficult. 
Intuitively, we could obtain all possible values in the domain of $\bm{\gamma}^{*}$ and numerically solve $\bm{\gamma}^{**}$ for each value and then save the result in a dictionary. In this way, we obtain a numerical representation of the function $h$ as a dictionary.
Then at inference time, for each value $\bm{\gamma}^{*}$ we could find the corresponding $\bm{\gamma}^{**}$ in the dictionary. 
Intuitively, the dictionary can be used for any unseen dataset as it is simply a different but equivalent representation to the analytical form.
Our ML-based solution can be seen as a less expensive approximation of such a dictionary-based approach. Instead of obtaining all possible values in the input domain, we pick a subset of random values (i.e., the training set), and instead of saving every pair of one-to-one mapping between $\bm{\gamma}^{*}$ and $\bm{\gamma}^{**}$, we compress the mapping by using an ML model. 
Similar to the dictionary-based approach, the trained ML model can be used for any dataset.
The way of learning an ML model to approximate functions (that are difficult to solve analytically) has also been adopted in other tasks recently~\cite{hruby2022learning, wu2022learning, wu2022learned, anonymous2023learning}. In these tasks, similar to ours, the trained model can be used for any unseen dataset~\cite{hruby2022learning, wu2022learning, wu2022learned, anonymous2023learning}. 
}

To manifest this idea, one challenge is that ML models require their inputs and outputs to have fixed dimensions, while the dimension of $\boldsymbol{\gamma}^*$ depends on the number of tuple pairs, which varies for different EM tasks.
Our approach is to first train a model to approximate $h$ for fixed-dimension $\boldsymbol{\gamma}^*$ (e.g., a 1024-dimensional $\bm{\gamma}$ is sufficient to represent at most 32 tuple pairs \revise{as there are $32\times 32=1024$ matching probabilities}). We then carefully decompose $\boldsymbol{\gamma}^*$ to sub-components with a maximum size of 1024 (i.e., clusters with 32 tuple pairs) and apply the model to each sub-component. 

\noindent\textbf{Training Data Generation.}
We randomly generate $10^5$ matching probability matrices of size $32\times 32$, each corresponding to a 1024-dimensional vector $\bm{\gamma}^*$. \revise{Each probability matrix corresponds to one training data point and we empirically observed that increasing the amount of training data from $10^5$ to $10^6$ does not bring meaningful improvement.}
 For each $\bm{\gamma}^*$, $\bm{\gamma}^{**}$ is obtained by Equation~\ref{eq:constrained_opt}. Specifically, by replacing $g(x^{(i,j)}, \theta)$ in $F(\theta, X, \boldsymbol{\gamma})$ with $\gamma^{*(i,j)}$:
\begin{small}
\begin{equation}
\begin{split}
\boldsymbol{\gamma}^{**}=\arg \min_{\boldsymbol{\gamma} \in Q}  \sum_{(i,j)} -\gamma^{(i,j)}\log\frac{\gamma^{*(i,j)}}{\gamma^{(i,j)}}
        -(1-\gamma^{(i,j)})\log\frac{1-\gamma^{*(i,j)}}{1-\gamma^{(i,j)}}\\
\end{split}
\end{equation}
\end{small}
We denote the big summation on the right-hand side as $h_1(\boldsymbol{\gamma}^{*},\boldsymbol{\gamma})$ so $\boldsymbol{\gamma}^{**}=\arg \min_{\boldsymbol{\gamma} \in Q} h_1(\boldsymbol{\gamma}^{*},\boldsymbol{\gamma})$.
We resort to expensive numerical optimizers to find $\boldsymbol{\gamma}^{**}$. Since this is done only once offline, we can afford expensive computation for a more accurate solution. 
Specifically, to account for the transitivity constraint $Q$, we add an additional transitivity loss to the objective function. The total amount of transitivity violations of all triplets of tuples is:
\begin{equation}
\small
l_{\text{transitivity}}(\bm{\gamma}) = \Sigma_{i,j,k}\text{Relu}(\gamma^{(i,j)}\gamma^{(i,k)}-\gamma^{(j,k)})
\end{equation}
The constrained solution $\boldsymbol{\gamma}^{**}$ can be obtained by minimizing the following expression with respect to $\bm{\gamma}$:
\begin{equation}
\small
\text{Loss}(\bm{\gamma}^*,\bm{\gamma}) = \alpha l_{\text{transitivity}}(\bm{\gamma}) + h_1(\bm{\gamma}^*,\bm{\gamma})
\end{equation}
where $\alpha$ is a hyperparameter controlling the preference between satisfying the transitivity constraint and minimizing the negative free energy function. We set $\alpha=100$ \revise{as we empirically observed that $\alpha=100$ ensures the transitivity constraint is satisfied in the final numerical solutions.} For each $\bm{\gamma}^*$, we numerically minimize $\text{Loss}(\bm{\gamma}^*,\bm{\gamma})$ with multiple optimizers including optimizers for nonconvex optimization~\cite{reddi2018adaptive,ma2020apollo} using an existing pytorch implementation~\cite{pytorch_opt}. Thus, for each $\bm{\gamma}^*$, we obtain multiple different solutions of $\bm{\gamma}^{**}$ where each solution is from a different optimizer. We then pick the solution with the smallest loss. Since we expect $\bm{\gamma}^{**}$ to be close to $\bm{\gamma}^{*}$, we always initialize $\bm{\gamma}^{**}$ as $\bm{\gamma}^{*}$ during optimization.

\noindent\textbf{Output Dimension Reduction.}
Predicting the $\bm{\gamma}^{**}$ vector requires the model to predict 1024 values at the same time, which would be very difficult. We instead reduce the task to predicting only a single value by exploiting a symmetry property of the task. 

\begin{figure}[htb!]
\vspace{-3mm}
  \centering
  \includegraphics[width=0.8\linewidth]{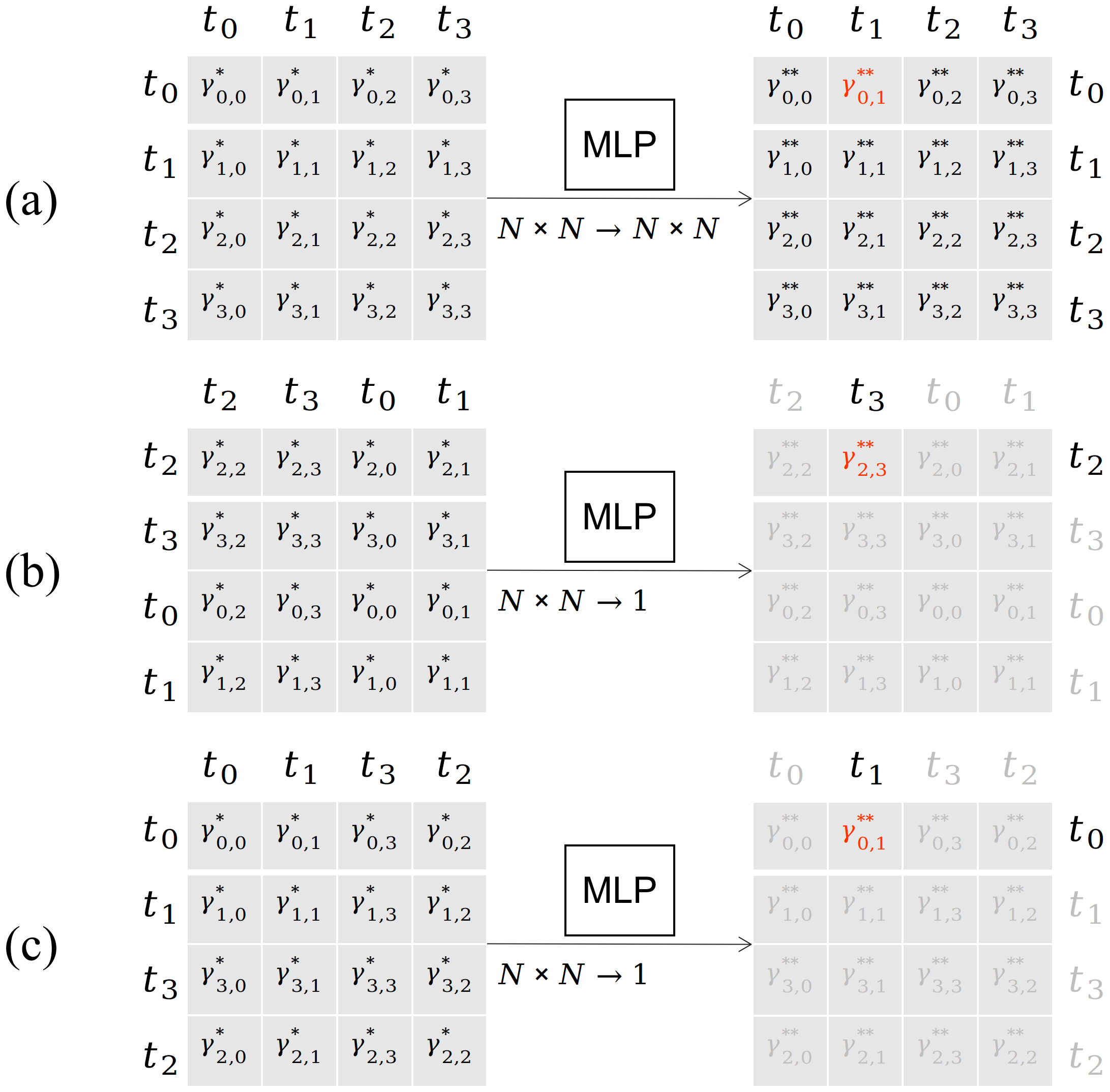}
    \vspace{-3mm}
  \caption{(a) Naive model. (b) Model with output dimension of 1. (c) Demonstration of invariance to permutation on tuples other than $t_0$ and $t_1$.}
  \label{fig:reduce_output}
\end{figure}

Consider the example shown in Figure~\ref{fig:reduce_output}. The naive approach of predicting all 1024 values is shown in Figure~\ref{fig:reduce_output}(a) where the input and output of the model are both a matrix of size $32\times 32$. Imagine we have a model learned to predict a single value in the red cell in Figure~\ref{fig:reduce_output}(a). Originally, the value in this cell is $\gamma_{0,1}^{**}$, which means the model predicts $\gamma_{0,1}^{**}$. Next in Figure~\ref{fig:reduce_output}(b), we swap $t_0$ with $t_{2}$ and swap $t_1$ with $t_3$. The rows and columns in the input matrix are swapped accordingly. Now the value in the red cell becomes $\gamma_{2,3}^{**}$ and the model predicts the value $\gamma_{2,3}^{**}$. Similarly, we can make the model predict \textit{any} value in the output matrix by swapping appropriate tuples, except the diagonal values which are known to be 1. 

Formally, let $h$ denote the model that takes an input matrix and predicts a single value in the red cell where the original value is $\gamma_{0,1}^{**}$. Let $S_{i,j}^{k,l}(\bm{\gamma}^*)$ denote swapping $t_k$ with $t_i$ and swapping $t_l$ with $t_j$. Then, $\forall i, j$ we have $\bm{\gamma}_{i,j}^{**}=h(S_{i,j}^{0,1}(\bm{\gamma}^*))$. In this way, we are able to predict any of the $32\times 32$ values in $\bm{\gamma}^{**}$ with a model $h$ that only predicts a single value. Therefore, we do not need to train a model that predicts the $32\times 32=1024$ values at the same time; we only need to train a model that predicts one value, which is much easier.

\noindent\textbf{Swapping-invariant Model Architecture.}
Consider that in Figure~\ref{fig:reduce_output}(c), when we swap $t_2$ with $t_{3}$, the model $h$ should still predict the value $\gamma_{0,1}^{**}$ and the swapping operation should not impact the prediction. 
This is a special swapping-invariance property of the problem which we also want to leverage.
In general, if we divide the tuples into groups $g_1 = \{t_0, t_1\}$ and $g_2=\{t_2,\dots\}$, then randomly swapping any tuples any number of times within each group should not change the prediction of $\gamma_{0,1}^{**}$, as the value in the red cell will always be $\gamma_{0,1}^{**}$ or $\gamma_{1,0}^{**}$, which are equivalent.  More formally, we want the following invariance to hold for $h$:
\begin{equation}
\label{eq:inv_property}
\small
\begin{split}
h(\bm{\gamma}^*) =& h(S_{0,1}^{1,0}(\bm{\gamma}^*))\\
h(\bm{\gamma}^*) =& h(S_{i,j}^{k,l}(\bm{\gamma}^*)), \forall i,j,k,l \text{ that } \{0,1\}\cap\{i,j,k,l\}=\emptyset 
\end{split}
\end{equation}
We would like to encode this invariance property directly into the model architecture. However, this is very challenging because when swapping the tuples, the corresponding rows and columns in the input matrix change at the same time. Let $\bm{\gamma}^*_{\text{mat}}$ denote the matrix form of $\bm{\gamma}^*$. The swapping operation $S_{i,j}^{k,l}(\bm{\gamma}^*)$ can be also written in a matrix form as $(P_{i}^{k}P_{j}^{l})\bm{\gamma}^*_{\text{mat}}(P_{i}^{k}P_{j}^{l})^T$ where $P_{i}^{k}$ is the permutation matrix~\cite{perm_mat} obtained by swapping the $i$th row and $k$th row of the identity matrix. Our core idea is to decompose the input matrix into an eigenvector matrix $V$ and eigenvalue matrix using the singular value decomposition $\bm{\gamma}^*_{\text{mat}}=VWV^T$. Subsequently,
\begin{equation}
\small
\begin{split}
(P_{i}^{k}P_{j}^{l})\bm{\gamma}^*_{\text{mat}}(P_{i}^{k}P_{j}^{l})^T =& (P_{i}^{k}P_{j}^{l})VWV^T(P_{i}^{k}P_{j}^{l})^T\\
=&(P_{i}^{k}P_{j}^{l}V)W(P_{i}^{k}P_{j}^{l}V)^T=V'W{V'}^T
\end{split}
\end{equation}
where $V' = P_{i}^{k}P_{j}^{l}V$. The $V'$ matrix is obtained by swapping the $i$th row with the $k$th row and swapping the $j$th row with the $l$th row in the matrix $V$. This means that swapping the rows and columns in $\bm{\gamma}^*_{\text{mat}}$ at the same time is equivalent to swapping only the rows in the eigenvector matrix $V$. Note that since the matrix $\bm{\gamma}^*_{\text{mat}}$ is symmetric, $W$ and $V$ are guaranteed to be real-valued. Also, while it might be tempting to use the Cholesky decomposition~\cite{Cholesky} instead (i.e., $\bm{\gamma}^*_{\text{mat}}=UU^T$), the problem is that $U$ can contain complex numbers which cannot be easily used as inputs for neural networks. 

To make the model satisfy the invariance in Equation~\ref{eq:inv_property}, we decompose the input $\bm{\gamma}^*_{\text{mat}}$ as $W$ and $V$, then use $W$ and $V$ as the new input to the model, so that we only need to ensure the model is invariant to row swapping operations on $V$. Specifically, similar to how we divide the tuples into two groups, we divide the rows in $V$ into two groups $g_1'=\{\text{1st row},\text{2nd row}\}$ and $g_2'=\{\text{3rd row}, \dots\}$, and we need to make sure that randomly swapping/shuffling any rows within each group does not change the model prediction. 

\begin{figure}[htb!]
\vspace{-4mm}
  \centering
  \includegraphics[width=\linewidth]{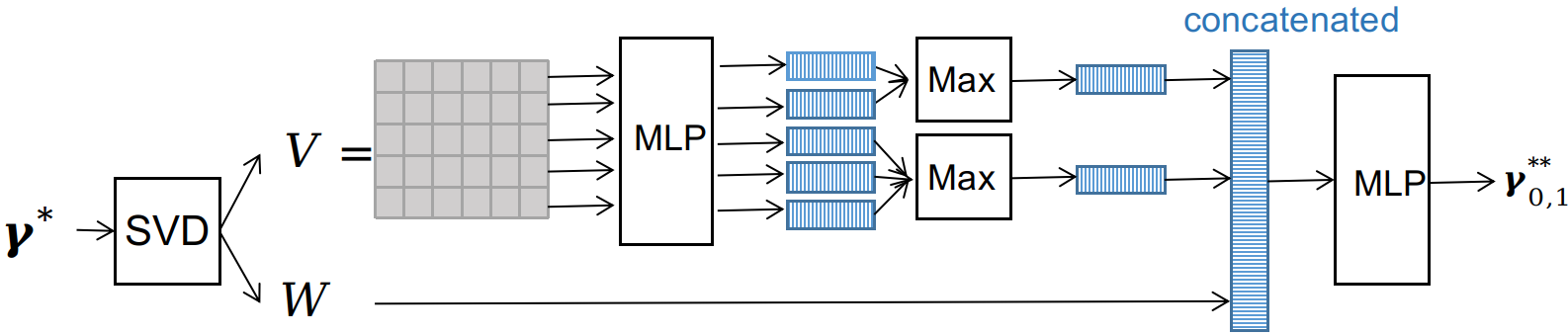}
    \vspace{-5mm}
  \caption{Transitivity model architecture.}
  \label{fig:model_arch}
     \vspace{-4mm}
\end{figure}

Inspired by PointNet~\cite{qi2017pointnet}, we present such a model architecture in Figure~\ref{fig:model_arch}. The input $\bm{\gamma^*}$ is first decomposed as $V$ and $W$. Each row in $V$ is then encoded by a neural network to obtain a row embedding vector. We subsequently take the maximum along every embedding dimension for embedding vectors in each group of rows ($g_1'$ and $g_2'$). This gives us one embedding vector for each group. Since the $\max$ operation is invariant to any swapping of elements within each group, the model architecture is invariant to row-swapping within each row group ($g_1'$ or $g_2'$) of $V$. The two embedding vectors of the two groups are then concatenated with the diagonal values in $W$ ($W$ is a diagonal matrix) to form one concatenated embedding vector. The concatenated vector finally passes through another neural network which predicts the value $\bm{\gamma}_{0,1}^{**}$.

\begin{figure}[htb!]
\vspace{-3mm}
  \centering
  \includegraphics[width=0.8\linewidth]{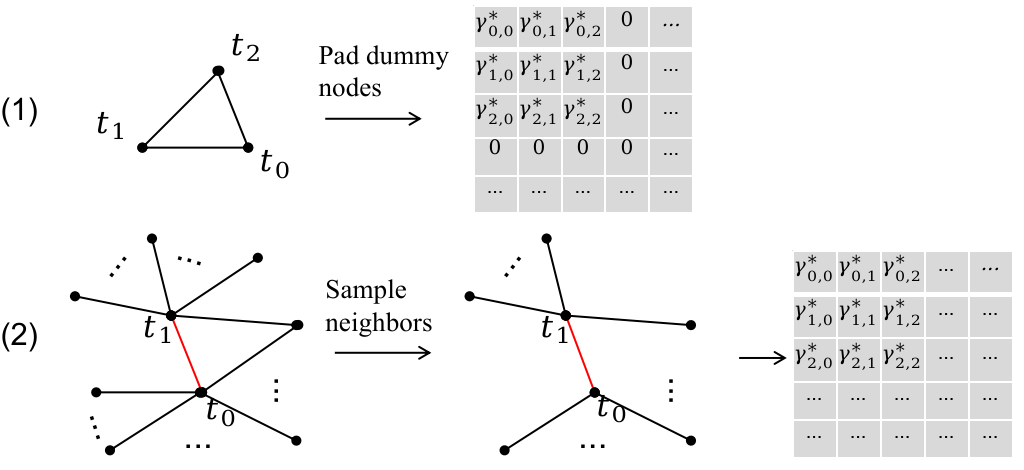}
    \vspace{-3mm}
  \caption{\revise{(1) A small connected component with a size $\leq$ 32. (2) A large connected component with a size $>$ 32. The matrix on the right is the input for the ML model.}}
  \label{fig:inference_arch}
     \vspace{-1mm}
\end{figure}

\noindent\textbf{Model Inference.} In online inference, we decompose a higher-dimensional input $\bm{\gamma^*}$ into many \textit{independent vectors} of 1024 dimensions, and we can apply $h$ on each vector. To do so, we form a graph $G_M$, where nodes are tuples and an edge exists between two nodes only if their matching probability exceeds 0.5. 
This is equivalent to a prior approach that relaxes the transitivity constraint $Q$ by only considering transitivity violations that involve at least two predicted matching pairs~\cite{ZeroER}. The intuition is that transitivity is useful only for tuple pairs predicted to be matches.  
After identifying the connected components of $G_M$, we can apply $h$ to tuple pairs in each connected component independently because there is no constraint that involves tuples from two different connected components (otherwise, the two components would have been merged). For each component, if it has fewer than 32 tuples (which covers most cases), we add dummy tuples to form a 1024-dimensional vector (or a 32$\times$ 32 matrix) \revise{as illustrated in Figure~\ref{fig:inference_arch}(1)}; if it has more than 32 tuples, then for each edge, we randomly sample 30 neighbors of the two nodes on the edge \revise{as illustrated in Figure~\ref{fig:inference_arch}(2)}; 
The random sampling is repeated ten times and we take the averaged prediction for that edge after applying $h$ to each sample. \revise{Each connected component with a size smaller than 32 will be processed by $h$ only once, so the involved edges will be processed only once.  
For a connected component with a size greater than 32, each edge will be processed 10 times as we take 10 random samples of its neighbors. Therefore, the time complexity is $O(N)$ where $N$ is the number of edges in the graph (which is the number of tuple pairs in the candidate set) and the time complexity of the overall SIMPLE-EM algorithm is still $O(M_IN\log(N))$. The space complexity of SIMPLE-EM is $O(N)$. }

\noindent\revise{\textbf{Computational complexity.} We summarize the computational complexity in all cases. For two-table EM, when one table is duplicate-free, the time complexity is $O(M_IN\log(N))$ where $M_I$ is the number of iterations (we empirically found $M_I<10$ suffices) and $N$ is the size of the candidate set; when two tables are duplicate-free, the time complexity is $O(M_IN(\log(N)+\min(N_l, N_r))$ where $N_l$ and $N_r$ are the number of tuples in the left and right table; when no table is duplicate-free, transitivity is not used and the time complexity is $O(M_IN\log(N))$. 
For single-table EM, the time complexity is $O(M_IN\log(N))$. In all cases, the space complexity is $O(N)$.
}

\vspace{-1mm}
\section{Experiments} \label{experiments-section}
We evaluate our proposed method along five dimensions:
\begin{itemize}[leftmargin=*]
\item \textit{Overall Performance.} How does the overall performance of our method (SIMPLE-EM) compare to other existing methods?
\item \textit{Transitivity.} How does our proposed method of handling transitivity compare to prior methods? %
\item \textit{Data shift.} How do traditional manual labeling and weak supervision behave in case of data shift? 
\item \textit{Sensitivity Analysis.} How sensitive is our method to different LFs?
\item \textit{Truth inference on general tasks.} How does the general form of our method (SIMPLE) work on general weak supervision tasks?
\end{itemize}

\vspace{-3mm}
\subsection{Experimental Setup}
\noindent\textbf{Hardware and Platform.} %
We use
a machine with a 2.20GHz Intel Xeon(R) Gold 5120 CPU, a K80 GPU and with 96GB 2666MHz RAM. \revise{In all experiments for all involved methods, we use GPU whenever possible and we use multi-core parallelization whenever possible.}

\noindent\textbf{Datasets.} We adopt commonly used real-world two-table and one-table benchmark datasets from the EM literature, such as those maintained by the Leipzig DB Group~\cite{kopcke2010evaluation, erhardws}, the Magellan project~\cite{magellandata} and the Alaska benchmark~\cite{crescenzi2021alaska}. The statistics of the datasets are shown in Table~\ref{tbl:datasets}. \revise{ The Monitor and Camera datasets are single-table datasets, and all other datasets are two-table datasets. Note that three datasets (IMDB-Rotten Tomatoes, Yellow Pages-Yelp, and Amazon-Barnes Noble) from the Magellan Repository~\cite{magellandata} only contain partial ground-truth (i.e., only a subset of matching pairs and a subset of non-matching pairs are provided). All other datasets have complete ground-truth. }
Also note that different records in the Alaska benchmark datasets (Monitor and Camera) can have different attributes~\cite{crescenzi2021alaska}; in these cases, we keep the 10 most common attributes and fill with NA for records missing these attributes. 
\revise{
We design a default blocking strategy for each dataset based on its most informative attributes (e.g., title) using the overlap of tokens. Specifically, we use the \textit{OverlapBlocker} from the py\_entitymatching package~\cite{overlap_blocker}; the candidate set size and blocking recall are shown in Table~\ref{tbl:datasets}. The same blocking is used for all baseline methods.}

\begin{table}[h]
\vspace{-3mm}
\caption{Benchmark EM datasets. \revise{
"-" for $N_{Non}$ denotes that all other tuple pairs are non-matches as the complete set of matches is provided for the dataset.}}
\vspace{-4mm}
\renewcommand{\arraystretch}{0.9}
\resizebox{\columnwidth}{!}{\begin{tabular}{|c|c|c|c|c|}
\hline
\textbf{Dataset} & \makecell{\textbf{\# tuples} \\L , R } &\makecell{\revise{\# matches $N_M$}\\ \revise{\# nonmatches $N_{Non}$}\\ \revise{\# unlabeled $N_U$} \\ \revise{\textbf{$N_M, N_{Non}, N_U$}}} & \textbf{\# attr} & \makecell{\revise{\textbf{candset}} \\ \revise{\textbf{size , recall}}} \\ \hline
Fodors-Zagats (FZ) & 533 ,  331               & \revise{112 , - , 0 }                 & 7       & \revise{2915 , 1.0}               \\ \hline
DBLP-ACM (DA) & 2616 , 2294            & \revise{2224 , - , 0}               & 4         &\revise{46456 , 0.998}              \\ \hline
DBLP-Scholar (DS) & 2616 , 64263           & \revise{5347 , - , 0}               & 4         &\revise{135327 , 0.913}               \\ \hline
Abt-Buy (AB)  & 1082 , 1093            & \revise{1098 , - , 0}               & 4           & \revise{164072 , 0.995}            \\ \hline
Amazon-Google (AG) & 1363 , 3226            & \revise{1300 , - , 0}               & 4       &\revise{42413 , 0.944}                 \\ \hline
\revise{Walmart-Amazon (WA)} & \revise{2554 ,  22074} & \revise{1154 , - , 0} & \revise{9}  &\revise{68265 , 0.879} \\ \hline
\revise{IMDB-Rotten Tomatoes (IR)} & \revise{2960 , 3093} & \revise{169 , 230 , 63399} & \revise{10}  &\revise{63798 , 1.0}  \\ \hline
\revise{Yellow Pages-Yelp (YY)}  & \revise{11840 , 5223} & \revise{131 , 271 , 4876} & \revise{6}  &\revise{5278 , 1.0}  \\ \hline
\revise{Amazon-Barnes Noble (ABN)}  & \revise{9836 , 9958} & \revise{233 , 143 , 67769} & \revise{11}  &\revise{68145 , 0.884}  \\ \hline
Monitor (M) & 16663 & \revise{26921 , - , 0} & 10  &\revise{99230 , 0.953} \\ \hline
Camera (C)  & 29788 & \revise{314315 , - , 0} &10   &\revise{1101318 , 0.979} \\ \hline
\end{tabular}}
\label{tbl:datasets}
\end{table}

\noindent\textbf{Algorithms Evaluated.} 
We first compare our SIMPLE-EM method to five state-of-the-art labeling models (or truth inference methods), which are selected based on prior benchmarks~\cite{zheng2017truth, zhang2021wrench}.
\vspace{-1mm}
\begin{itemize}[wide,labelindent=0pt]
\item \textit{Majority Vote (MV):} For each tuple pair, the labeling model's prediction is the most common label given by the labeling functions.

\item \textit{Dawid and Skene's Method (D\&S):} This method models each LF's confusion matrix with respect to the ground-truth and optimizes the parameters with the Expectation-Maximization algorithm~\cite{dawid1979maximum}. 

\item \textit{Enhanced Bayesian Classifier Combination (EBCC):} This method models the joint distribution of LFs with matrix decomposition to reduce the number of parameters~\cite{2019-ICML-Exploiting-Worker-Correlation-for-Label-Aggregation.pdf}. This is the state-of-the-art variant of the Bayesian Classifier Combination based methods. 

\item \textit{Snorkel (SN):} This is the labeling model used by the Snorkel system~\cite{2018-VLDB-Snorkel-System.pdf}. We use the latest open-source implementation~\cite{github_snorkel} which performs truth inference with a matrix completion model~\cite{ratner2019training}. %

\item \textit{Flying Squid (FS):} This is an efficient alternative to the PGM model. It provides a closed-form solution for a triangulated PGM under some assumptions, obviating the need for an iterative EM algorithm or stochastic gradient descent~\cite{fu2020fast}.
\vspace{-1mm}
\end{itemize}

We further compare to \revise{EM solutions designed to require no or less label data including} the state-of-the-art unsupervised EM solution, \revise{one pre-trained language model based solution}, and two active learning based solutions:
\vspace{-1mm}
\begin{itemize}[wide,labelindent=0pt]
\item \textit{ZeroER (ZE):} This is the state-of-the-art unsupervised entity matching solution~\cite{ZeroER}. ZeroER tailors the Gaussian Mixture Model to EM by considering several EM-specific properties. 
ZeroER uses Magellan~\cite{2016-VLDB-Magellan.pdf} to do feature engineering.
\revise{We use the official implementation~\cite{zeroer_github} that supports both two-table and one-table EM.}
\item \revise{\textit{Ditto:} This is the state-of-the-art entity matching system based on pre-trained language models~\cite{li2020deep}. Ditto casts the entity matching task as a sequence pair classification task and works by fine-tuning a pre-trained language model for the task. Since the model can capture various semantic meanings of textual data, Ditto is expected to require fewer labeled examples.} 
\item \textit{Active learning based Random forest (AL-RF):} We use the implementation from the most widely used active learning library modAL~\cite{modAL2018} and use the default query strategy (uncertainty sampling). Active learning requires feature engineering, so we use Magellan~\cite{konda2018magellan} to automatically generate features. We use the default strategy
to handle class imbalance. 
\item \textit{Active learning based Random forest with SMOTE (AL-RF-S):} This is a variant of AL-RF that uses SMOTE~\cite{chawla2002smote} to handle class imbalance. 
\end{itemize}

\noindent\textbf{Setups for Various Algorithms.}
The candidate set size after blocking on Monitor and Camera is still huge, so we take a subsample on these two datasets. To not lose any matches in the candidate set, we sample by keeping the tuple pairs where at least one of the two tuples has matches. The candidate set size and recall in Table~\ref{tbl:datasets} are obtained after subsampling. 
For all methods that need a seed, the reported results are the average results of five runs. For methods (e.g., Snorkel) that require a class weight parameter to handle class imbalance, we obtain the class weights by counting the number of matches and non-matches from the Majority Vote method. \revise{For active learning and Ditto, we evaluate the performance on a hold out test set which includes 20\% of the data.}

\noindent\textbf{Performance Metric.} We use F1 score as our performance metric since EM is a binary classification task with unbalanced classes.

\vspace{-2mm}
\subsection{Labeling Function Development}
\label{ssec:lf_dev}
Though we consider LF development as an orthogonal task and focus on accurately combining a given set of LFs, we provide the details of how we obtain the LFs for EM datasets as there is no existing open-source LFs.
We use an existing tool, the Panda IDE~\cite{panda_demo}, that provides a visual interface to develop LFs efficiently for each dataset. 
Our developed LFs are available at~\cite{SIMPLE_code}.%

We developed LFs for all datasets in Table~\ref{tbl:datasets} in the following order: FZ, DA, DS, AB, AG, M, C, \revise{WA, IR, YY, and ABN}. When writing LFs for one dataset, we may reuse LFs from different attributes of the same dataset or LFs from a prior dataset. To properly measure the effort spent, we report the number of new LFs that require significant effort on each dataset. Specifically, an LF that requires more than 15 seconds is counted as a new LF, while an LF that is obtained in less than 15 seconds by simply modifying the attribute or distance function of an existing LF does not count as a new LF. \revise{On each dataset, the number of new LFs is smaller than the total number of LFs because we reuse LFs from prior datasets or across multiple attributes of the same dataset and these "duplicate" LFs are not counted as new LFs. }

The statistics of new LFs and total time spent on all datasets are shown in Table~\ref{tbl:lf_time_spent}. The number of LFs ranges from 8 to 16 for the datasets. However, most of the LFs are obtained rapidly by changing the attribute or distance function of existing LFs. Typically, one only needs to develop a few new LFs with some actual effort for a new dataset; the required time ranges from 10 to 50 minutes.

\begin{table}[h]
\vspace{-3mm}
\caption{Time spent for developing LFs }
\vspace{-3mm}
\renewcommand{\arraystretch}{0.9}
\resizebox{1\columnwidth}{!}{
\begin{tabular}{|c|c|c|c|c|c|c|c|c|c|c|c|}
\hline
     & FZ & DA& DS & AB &AG&\revise{WA}&\revise{IR}&\revise{YY}&\revise{ABN} &M &C \\ \hline
\# of LFs& 12 & 13& 16 & 13&13 & \revise{14} &\revise{8} &\revise{10} &\revise{13}&10&11 \\ \hline
\# of new LFs& 5 &  6 & 4 & 4 &3&\revise{2}&\revise{1}&\revise{1} &\revise{2} & 3 & 2\\ \hline
time spent, minutes  & 30     &50 & 45    &30 & 20 & \revise{15} &\revise{10} &\revise{10}&\revise{15} &25 & 15      \\ \hline
\end{tabular}}
\label{tbl:lf_time_spent}
\end{table}

\subsection{Overall Performance}
\label{ssec:overall_perf}
\noindent\textbf{Labeling Performance.} The performance results for weak/un-supervised  methods are shown in Table~\ref{tbl:performance}.
Our method outperforms all baseline methods significantly, achieving the highest F1 score on 9 out of the 11 datasets, and only under-performing EBCC on the Camera (C) dataset and D\&S on the IR dataset by less than 1\%. 
Snorkel is the best-performing weak supervision baseline on most datasets, and our model achieves 10\% higher F1 score on average across all datasets. %
\revise{Snorkel performs poorly on the AG dataset; one possible reason for this is that the assumptions Snorkel makes (e.g., LFs are conditionally independent) are violated on AG. All methods perform badly on the WA dataset as the dataset is very dirty.}

The unsupervised method ZeroER does not work as well as weak supervision methods. This is because unsupervised methods do not use any supervision signals, so they typically only work well on simple datasets like Fodors-Zagats (FZ). 

\begin{table}[h]
\vspace{-3mm}
\caption{F-1 scores for weak/un-supervsied methods}%
\setlength{\tabcolsep}{2.5pt}
\vspace{-3mm}
\renewcommand{\arraystretch}{0.9}
\resizebox{0.8\columnwidth}{!}{\begin{tabular}{|l|l|lllll|l|}
\hline
        &   & \multicolumn{5}{c|}{Weak Supervision}& Unsupervised  \\ \hline
        & SIMPLE-EM & MV    &D\&S   & EBCC        & FS        & SN  &ZE           \\ \hline
FZ      & \textbf{0.996}            & 0.848  & 0.973        & 0.978          & 0.644          &           0.942&  0.992        \\ \hline
DA      & \textbf{0.991}              & 0.726   &0.339        &  0.238          & 0.324          &        0.958& 0.957         \\ \hline
DS      &\textbf{0.911}              &  0.908  & 0.896      & 0.824          & 0.421          &            0.904& 0.863          \\ \hline
AB      &\textbf{0.906}       &  0.628  &0.686       & 0.327          &0.689           &     0.776& 0.520            \\ \hline
AG      & \textbf{0.555}         & 0.439    & 0.426     &  0.369        &0.217           &        0.199 & 0.484            \\ \hline
\revise{WA} & \revise{\textbf{0.499}} & \revise{0.397} & \revise{0.332} & \revise{0.395} &\revise{0.085} & \revise{0.363} & \revise{0.400}\\ \hline
\revise{IR} & \revise{0.989} & \revise{0.985} & \revise{\textbf{0.997}} & \revise{0.911} &\revise{0.982} & \revise{0.958} & \revise{0.968}\\ \hline
\revise{YY} & \revise{\textbf{0.969}} & \revise{0.968} & \revise{0.952} & \revise{0.960} &\revise{0.956} & \revise{0.953} & \revise{0.684}\\ \hline
\revise{ABN} & \revise{\textbf{0.897}} & \revise{0.834} & \revise{0.792} & \revise{0.821} &\revise{0.240} & \revise{0.809} & \revise{0.839}\\ \hline
M       & \textbf{0.887}    & 0.780    & 0.708    &  0.737         & 0.665          &       0.812    & 0.325        \\ \hline
C     &   0.872      & 0.791    & 0.791     &   \textbf{0.884}        & 0.865          &       0.817  &   0.477       \\ \hline
\revise{Avg.} &   \revise{\textbf{0.861}}      &  \revise{0.755}    &\revise{0.717}      &   \revise{0.677}        & \revise{0.553}          &     \revise{0.772}  &  \revise{0.683}       \\ \hline
\end{tabular}}
\label{tbl:performance}
\end{table}

\noindent\revise{\textbf{Comparison to Ditto.} We compare our method to Ditto, a method based on pre-trained language models that is expected to require less data. We use the implementation from the official GitHub repository and use the default configurations~\cite{ditto_github}. For some datasets, the official repository also provides pre-split training, validation and test sets that are small subsamples of the candidate set~\cite{li2020deep}. We first confirm that we are able to get comparable results with the original paper using the provided training, validation and test sets. However, the provided training, validation and test sets are only small subsamples of the candidate set; for example, the provided Walmart-Amazon dataset only includes about 10000 pairs, which is just 15\% of our candidate set size.
Therefore, to evaluate the performance of Ditto in a more realistic setting and also to ensure the setting of Ditto is comparable to our method, in our experiment we use all data in the candidate set (with ground-truth labels) and randomly split the data into training, validation, and test sets by a ratio of 3:1:1. We report the results in Table~\ref{tbl:ditto}. 
}
\begin{table}[h]
\vspace{-3mm}
\caption{\revise{Comparison to Ditto (F-1 score)}}
\vspace{-3mm}
\setlength{\tabcolsep}{1.5pt}
\renewcommand{\arraystretch}{0.8}
\resizebox{1\columnwidth}{!}{
{\color{revision}
\begin{tabular}{|c|c|c|c|c|c|c|c|c|c|c|c|}
\hline
     & FZ & DA& DS & AB &AG&WA&IR&YY&ABN &M &C \\ \hline
SIMPLE-EM& 0.996 & 0.991& 0.911 & 0.906&0.555 & 0.499 &0.989 &0.969 &0.897&0.887&0.872 \\ \hline
Ditto& 0.951 &  0.967 & 0.933 & 0.283 &0.275 &0.262&0.716&0.861 &0.708 & 0.844 & 0.627\\ \hline
\end{tabular}}}
\label{tbl:ditto}
\vspace{-4mm}
\end{table}

\revise{
Ditto is better than our method only on the Dblp-Scholar (DS) dataset. Ditto likely falls short of our method for the following reasons: 
(1) Ditto is sensitive to the creation of training/validation/test sets. For example, on the Abt-Buy dataset, Ditto gets a F1 score of 0.821 on the provided training/validation/test sets in the GitHub repository but only gets an F1 score of around 0.283 on random splits (we got similar results on several splits).  
(2) Since Ditto is based on pretrained language models, Ditto is expected to have advantages on datasets with many text attributes and may not work well on datasets with numerical features and categorical features. 
}

\noindent\textbf{Comparison to Active Learning.}
Active learning is also a common technique to obtain labeled data. We compare our method to two active learning methods (AL-RF and AL-RF-S). The results are shown in Table~\ref{tbl:performance_al}. Note that in Table~\ref{tbl:performance_al}, for each dataset, we report the best result between the two active learning methods (the performance of the two methods is similar across datasets, so we do not report them separately). Human time is estimated by assuming each label takes three seconds.
 On six out of eight datasets, active learning is not able to match the performance of our method even after querying for labels on all data points \revise{(which is equivalent to a random forest classifier trained using all labels)}. 
Even on the three datasets (DS, WA and C) where active learning matches the performance of our method, active learning requires several hundreds or thousands of labels. The required human time can be as much as 200 minutes, which is significantly more than the required human time for our method (shown in Table~\ref{tbl:lf_time_spent}).
 
 Active learning likely does not work as well as our method for the following reasons: (1) Weak supervision holistically incorporates (weak) information of all data points to infer the ground-truth labels, while active learning infers the decision boundary based on the set of most uncertain data points selected on its query strategy which might not be reliable.
Furthermore, on difficult datasets where the decision boundary is complicated, active learning would still need to select many data points. (2) Our method considers the transitivity property of EM which provides additional signals.

\begin{table}[h]
\vspace{-3mm}
\caption{Comparison to active learning. Note we report the best result of the two active learning methods (AL-RF and AL-RF-S) on each dataset. "-" denotes that active learning is not able to match our method's performance. Human time is estimated by assuming each label takes three seconds. \revise{Note that the three datasets (IR, YY, ABN) with only partial ground-truth are not included as active learning may query data points not in the ground-truth.}}
\setlength{\tabcolsep}{1pt}
\vspace{-3mm}
\renewcommand{\arraystretch}{0.9}
\resizebox{\columnwidth}{!}{\begin{tabular}{|l|l|lll|ll|}
\hline
        &   & \multicolumn{3}{c|}{AL matches SIMPLE}& \multicolumn{2}{c|}{AL queries all labels}  \\ \hline
        & SIMPLE-EM & \# of labels     &\% of labels &human time (min)   & F1        &  \# of labels         \\ \hline
FZ      & 0.996              & -  & -   &-     & 0.985          & 2332           \\ \hline
DA      & 0.991              & -  &-   &-     &  0.981          & 37165             \\ \hline
DS      &0.911              &  460  & 0.4\% &23      & 0.938          & 108262     \\ \hline
AB      & 0.906       &  -  &-   &-    & 0.510          &131258     \\ \hline
AG      & 0.555         & -   & -  &-   &  0.539        &33931      \\ \hline
\revise{WA} & \revise{0.499} & \revise{350} & \revise{0.5\%} &\revise{17.5} & \revise{0.695}&\revise{ 3150} \\ \hline
M       & 0.887     & -     & - &-   &  0.848         & 79384                       \\ \hline
C     &  0.872     & 4310    & 0.5\%  &215   &   0.949        & 881055              \\ \hline
\end{tabular}}
\label{tbl:performance_al}
\vspace{-1mm}
\end{table}

\noindent\textbf{\revise{Running Time.}}
\revise{
The memory requirements of the methods are comparable except that Majority Vote requires much less memory and Ditto requires much more memory.
The running times of all methods are shown Table~\ref{tbl:runing_time}. 
Note the table only shows machine time, and human time for labeling is not included. For weak supervision methods, the running time includes the time for applying the LFs to all tuple pairs in the candidate set to obtain the weak labels and the time for inferring the ground-truth labels. The unsupervised method ZeroER (ZE) and active learning method (AL-RF) require feature engineering to be done. We use Magellan~\cite{konda2018magellan} to automatically perform feature engineering for the two methods following prior work~\cite{ZeroER}. For these two methods (ZE and AL-RF), the running time includes the time for feature engineering and the time for training and inference. For Ditto, we use the default configuration from the official implementation~\cite{ditto_github} and the running time includes the time for preprocessing the textual data (e.g., tokenization), the time for data augmentation, the time for text summarization, and the time for training and inference~\cite{li2020deep}.
}

\revise{
The running time of SIMPLE-EM is greater than other weak supervision methods, as performing cross validation to select the hyperparameters for the random forest classifier can be relatively expensive. 
However, this can be alleviated by using more CPUs because cross validation can be easily parallelized. Overall, the unsupervised method ZeroER and the active learning method AL-RF have much higher running time than SIMPLE-EM because these two methods require feature engineering, which is expensive. In addition, AL-RF updates the model when querying each new data point which is expensive.
 Ditto has the longest running time as it involves multiple expensive steps like tokenizing the text data, performing data augmentation and text summarization, and training a deep learning model. Note that for the three datasets (IR, YY and ABN) with partial labeled data, we only use the small labeled subset of data for training Ditto, so the running time of Ditto is extremely small on these three datasets.
}

\begin{table}[h]
\vspace{-3mm}
\caption{\revise{Running time (minutes) for all methods. Note only machine time is included even for active learning (AL-RF).}}
\vspace{-3mm}
\color{revision}
\renewcommand{\arraystretch}{0.9}
\resizebox{\columnwidth}{!}{\begin{tabular}{|l|l|lllll|l|ll|}
\hline
        &    \multicolumn{6}{c|}{Weak Supervision}& Unsupervised & \multicolumn{2}{c|}{Supervised} \\ \hline
        & SIMPLE-EM & MV    &D\&S   & EBCC        & FS        & SN  &ZE  &AL-RF & Ditto         \\ \hline
FZ      & 0.6            & 0.1  & 0.2 & 0.1          & 0.1          &           0.1 &  1.5    &3.7 & 4.4   \\ \hline
DA      & 3.7              & 0.6   &1.8        &  1.2          & 0.7          &       0.6 & 16.3    &18.1 & 59.2    \\ \hline
DS      &15.3             &  3.1  & 7.7      & 3.5          & 3.1          &            3.3 & 52.4      &57.2  &  113.7   \\ \hline
AB      &17.9       &  3.7  &8.4       & 4.7          &3.9 &     3.7& 71.8      &81.5 & 140.3     \\ \hline
AG      & 7.9         & 0.8    & 3.3     &  1.1        &0.9           &        0.8 &22.4     &26.9 &127.4        \\ \hline
WA & 6.8 & 1.4 & 3.6 & 1.7 &1.5 & 1.4& 179.2 & 194.1&214.8 \\ \hline
IR & 6.5 & 0.9 & 2.7 & 1.3 &1.1 & 0.9 & 28.2 &- &1.4 \\ \hline
YY & 1.7 & 0.3 & 0.4 & 0.3 &0.3 & 0.3 & 4.5 &- &0.8 \\ \hline
ABN & 12.8 & 0.6 & 11.6 & 1.1 &0.9 & 0.6 & 30.3 &- &1.2 \\ \hline
M       & 5.5    & 0.7   & 5.1    & 0.9         & 0.8         &       0.7    & 128.1  &133.4 & 151.3      \\ \hline
C     &   46.5      & 6.3    & 42.2     &   15.1       & 7.4 &       6.3  &  197.9  &216.0 &248.7     \\ \hline
Avg. &   11.4      & 1.7    &7.9      &   2.9       & 1.9          &     1.7  &  66.6  &91.4 &96.7     \\ \hline
\end{tabular}}
\label{tbl:runing_time}
\end{table}

\noindent\textbf{End Model EM Performance.} Our generated labels can be used to train any downstream models, such as existing supervised ML methods. We use DeepMatcher~\cite{2018-SIGMOD-DeepMatcher-Design-Space-Exploration.pdf} 
as an example downstream model to demonstrate the effectiveness our generated labels for training an end model. Using the open-source implementation~\cite{github_deepmatcher}, we compare the DeepMatcher model trained with our generated labels with the DeepMatcher model trained with ground-truth labels.
For each dataset, we divide the tuple pairs after blocking into training, validation, and test sets by a ratio of 3:1:1. \textbf{(1) SIMPLE-EM Labels:} We use our generated labels for the training and validation set to train a model and use the ground-truth labels for the test set to evaluate the trained model. \textbf{(2) Ground-Truth Labels:} We train another end model with ground-truth labels in the training and validation set. 
To measure the labeling effort saved by SIMPLE-EM, we gradually increase the training set size to match the performance of the end model trained on our generated labels. %
We report the number of ground-truth labels required to match the performance of the end model trained on our labels
and the number of ground-truth labels (when F1 score is converged) where adding more ground-truth labels does not improve F1 score by a meaningful amount.

\begin{table}[h]
\vspace{-3mm}
\caption{DeepMatcher trained on SIMPLE-EM labels vs DeepMatcher trained on Ground-Truth (GT) labels. \revise{The "*" symbol on the converged \# GT labels denotes that the dataset only has partial ground-truth labels and the number precedes "*" is 80\% of the ground-truth labels (the other 20\% are used as the test set).}}
\setlength{\tabcolsep}{1pt}
\vspace{-4mm}
\renewcommand{\arraystretch}{0.9}
\resizebox{\columnwidth}{!}{\begin{tabular}{|c|c|c|c|c|c|c|c|c|c|c|c|c|}
\hline
     & FZ & DA& DS & AB &AG&\revise{WA}&\revise{IR}&\revise{YY}&\revise{ABN}&M &C &Avg. \\ \hline
\begin{tabular}{c}F1 of DeepMatcher\\on SIMPLE-EM labels \end{tabular}    & 0.979     &0.978 & 0.926                   &0.673 & 0.741 &\revise{0.450} &\revise{1.0} &\revise{1.0} &\revise{0.898} &0.956 & 0.929&0.866      \\ \hline
\begin{tabular}{c}\# GT labels to\\ match above F1 \end{tabular}    & 2333     &37170 & 48284                   &164078 & 11015 &\revise{10947} &\revise{-} &\revise{-} &\revise{265} &7586 & 8866&32282      \\ \hline
\begin{tabular}{c}Converged F1 \\ \# GT labels \end{tabular} &\begin{tabular}{c}0.979\\2333 \end{tabular}      &\begin{tabular}{c}0.978\\37170 \end{tabular}                 & \begin{tabular}{c}0.956 \\120710 \end{tabular} &\begin{tabular}{c}0.692 \\218770 \end{tabular}  &\begin{tabular}{c}0.845 \\55075 \end{tabular}&\begin{tabular}{c}\revise{0.631} \\ \revise{16421} \end{tabular} &\begin{tabular}{c}\revise{0.964} \\\revise{320$^*$} \end{tabular} &\begin{tabular}{c}\revise{0.923} \\ \revise{322$^*$} \end{tabular} & \begin{tabular}{c}\revise{0.903} \\\revise{285$^*$} \end{tabular}&\begin{tabular}{c}0.999 \\75860 \end{tabular} &\begin{tabular}{c}0.999 \\122200 \end{tabular}&\begin{tabular}{c}0.897 \\59042 \end{tabular}\\ \hline
\end{tabular}}
\label{tbl:end_model}
\end{table}

Overall, when using ground-truth labels, at least several thousand labels are required for the end model to match its performance when trained on the generated labels from SIMPLE-EM. In addition, the F1 score of end model trained on SIMPLE-EM labels is comparable (though on average 3.1\% worse) to the converged F1 of the end model trained on sufficient amount of ground-truth labels. 

We observed that the performance of the end model trained on SIMPLE-EM's labels can sometimes be better than the original labeling performance. 
On the AG dataset in particular, the end model F1 score is $0.741$, which is about 20\% better than the original labels ($0.555$).
The reason is that the end model can incorporate additional information (e.g., textual features) that is different from the information used in LFs. This phenomenon is also observed in multiple prior works on weak supervision~\cite{varma2018snuba, das2020goggles, wu2022cluster}.

On the AG dataset, the end model trained with ground-truth labels (which have F1 score of $1$) gets an F1 score of at most $0.845$, which is only $10\%$ better than the end model trained on SIMPLE-EM labels. This is in spite of the fact that the SIMPLE-EM labels have a $50\%$ worse F1 score than the ground-truth labels. Our takeaway is that training end models with noisy and weak supervision often suffices, while accurate supervision by an expensive process of labeling individual data points may not be necessary. This confirms prior empirical findings~\cite{snubaVLDB18, 2018-VLDB-Snorkel-System.pdf} and theoretical derivations~\cite{robinson2020strength, ratner2016data}.

\vspace{-2mm}
\subsection{Handling Transitivity Constraint} \label{ssec:ablation-study}
We conduct experiments to compare different ways of handling transitivity. We compare the following methods:
\begin{itemize}[wide,labelindent=0pt]
\vspace{-1mm}
\item \textit{No trans:} This is to ignore transitivity and directly use our base labeling model (SIMPLE) in Section~\ref{sec:proposed_truth_inference}.
\item \textit{SIMPLE-EM:}
 This is our proposed method in Section~\ref{tup-trans-section}.
 \item \textit{ZeroER Trans:} We replace our component of handling transitivity with the one in ZeroER~\cite{ZeroER}. \revise{ZeroER uses a projection based heuristic to solve Equation~\ref{eq:constrained_opt} to enforce transitivity.}
 \item \textit{Postprocessing:} This is the traditional way of  handling transitivity in a postprocessing step.  \revise{Specifically, the \textit{No trans} method (SIMPLE) is used to obtain a prediction, then a postprocessing step is done on the prediction to enforce transitivity}. The way to perform postprocessing differs for two-table datasets and single-table datasets. (1) On single-table datasets, we have the matching probability for all pairs, so it is possible to use clustering methods in ER literature. We use the method adopted by dedupe~\cite{gregg2015dedupe} (hierarchical clustering with centroid linkage~\cite{murtagh2012algorithms}) and we also use the implementation in dedupe~\cite{gregg2015dedupe}.
(2) On two-table datasets, we do not have the matching probabilities for tuple pairs from the same table. In order to do postprocessing, following prior work~\cite{ZeroER}, we assume that the left and right table are duplicate-free and the matching probability of any tuple pair from the same table to be 0 for all two-table datasets. 
 For example, we have three tuples $t_{l_1}$, $t_{l_2}$, and $t_{r_1}$ where $t_{l_1}$ and $t_{l_2}$ come from the left table and $t_{r_1}$ is from the right table. We assume the matching probability of $t_{l_1}$ and $t_{l_2}$ to be 0, i.e., $\gamma^{(l_1, l_2)}=0$. When the transitivity constraint is violated, we keep the cross-table tuple pair with higher probability as a match. For example, when $\gamma^{(l_1, r_1)}=0.8$ and $\gamma^{(l_2, r_1)}=0.9$, the transitivity constraint is violated and we keep $(t_{l_2}, t_{r_1})$ as a match and discard the tuple pair $(t_{l_1}, t_{r_1})$. 

\end{itemize}

\begin{table}[h]
\vspace{-3mm}
\caption{Different methods to handle transitivity.}
\vspace{-3mm}
\setlength{\tabcolsep}{1pt}
\renewcommand{\arraystretch}{0.85}
\resizebox{\columnwidth}{!}{
\begin{tabular}{|c|c|c|c|c|c|c|c|c|c|c|c|c|}
\hline
     & FZ & DA& DS & AB &AG &\revise{WA}&\revise{IR}&\revise{YY}&\revise{ABN} &M &C&Avg. \\ \hline
No Trans&0.978 &  0.765 & \textbf{0.911} & 0.697 &\textbf{0.555} &\revise{0.493}&\revise{\textbf{0.988}} &\revise{\textbf{0.969}}&\revise{0.884}& 0.781& 0.832&\revise{0.805}\\ \hline
SIMPLE-EM& \textbf{0.996} & \textbf{0.991}& \textbf{0.911} & \textbf{0.906}&\textbf{0.555}&\revise{\textbf{0.499}}&\revise{\textbf{0.989}} &\revise{\textbf{0.969}}&\revise{\textbf{0.897}}&\textbf{0.887}&\textbf{0.872}&\revise{\textbf{0.861}} \\ \hline
ZeroER Trans  & 0.993     &\textbf{0.991} & 0.880    &0.794 & 0.413 &\revise{\textbf{0.499}}&\revise{0.988} &\revise{0.670}&\revise{\textbf{0.897}}&0.694 & 0.142&\revise{0.724   }  \\ \hline
Postprocess  & 0.990     &0.979 & 0.625    &0.343 & 0.486 &\revise{0.494}&\revise{0.985} &\revise{0.670}&\revise{0.894}&0.610 & 0.633 &\revise{0.701}     \\ \hline
\end{tabular}}
\label{tbl:transitivity_handling}
\vspace{-1mm}
\end{table}

The results are shown in Table~\ref{tbl:transitivity_handling}. Overall, our method of handling transitivity works the best on all datasets and improves F1 score by about 9\% on average. The method of handling transitivity in ZeroER is not robust across datasets as it employs a greedy algorithm to correct each triplets of tuple pairs that violates transitivity~\cite{ZeroER}. In contrast, our method holistically considers all tuple pairs when enforcing transitivity.  Postprocessing also does not work well \revise{and is even worse than \textit{No trans}, e.g. on AB}. \revise{This is because postprocessing can introduce spurious matches or remove true matches~\cite{baas2021exploiting, ZeroER}.} 
Intuitively, postprocessing separates the clustering stage from the inference stage and therefore each stage uses less information, while our method considers clustering as an integral part of the inference stage and thus exploits all information holistically.

\subsection{\revise{Controlled Study of Transitivity}}
\label{ssec:study_trans}

\subsubsection{\revise{A Survey on Transitivity Violations in real-world datasets}}
\label{sssec:survey_trans}
\revise{
We investigate the frequency of transitivity violations in the ground-truth of real-world datasets and examine the cause of the violations.
}

\revise{
 We first introduce how we detect transitivity violations in the ground-truth. For single-table datasets, three tuples $t_i$, $t_j$ and $t_k$ constitute a violation if in the ground-truth $(t_i, t_j)$ and $(t_i, t_k)$ are matches but $(t_j, t_k)$ is a non-match. For two-table datasets, since only ground-truth labels of the cross-table tuple pairs are provided~\cite{magellandata}, we can only identify transitivity violations with the following method: Let $t_{l,i}$ and $t_{l,j}$  denote two tuples from the left table; Let $t_{r,i}$ and $t_{r,j}$  denote two tuples from the right table; The transitivity property is violated when $(t_{l,i}, t_{r,i})$, $(t_{l,j}, t_{r,i})$ and $(t_{l,i}, t_{r,j})$ are matches and $(t_{l,j}, t_{r,j})$ is not a match. To see this, when $(t_{l,i}, t_{r,i})$ and $(t_{l,j}, t_{r,i})$ are matches, the tuples $t_{l,i}$ and $t_{l,j}$ are the same entity by the transitivity property, so if $(t_{l,i}, t_{r,j})$ is a match $(t_{l,j}, t_{r,j})$ must be a match, otherwise the transitivity property is violated.}

\revise{
In the 11 datasets we used in our experiments, only two datasets (DS and WA) contain transitivity violations in the ground-truth (on which SIMPLE-EM still outperforms other methods, see Table~\ref{tbl:performance}.). %
We further inspected all 30 datasets in the Magellan repo~\cite{magellandata}, and found that only 6 datasets contain transitivity violations in the ground. For these 6 datasets, on average only 4\% of the labeled tuple pairs in the ground-truth are involved in a violation. This validates a belief commonly-held in the literature that transitivity is mostly satisfied in real-world scenarios~\cite{bohm2012linda, lacoste2013sigma, ZeroER}.}

\revise{To investigate the causes of transitivity violations, we manually inspect a random sample of the detected violations. We found that in 40\% of the cases, the violation of transitivity is due to incorrect matching pairs in the ground-truth.
For example, $(t_i, t_j)$ is a match and $(t_i, t_k)$ should not be a match, but the ground-truth includes both $(t_i, t_j)$ and $(t_i, t_k)$ as matches. 
The other 60\% of the cases are caused by incomplete matching labels in the ground-truth. For example, both $(t_i, t_j)$ and $(t_i, t_k)$ are actual matches and are included in the ground-truth, $(t_j, t_k)$ is also a match (based on a manual inspection) but is not included as a match in the ground-truth. 
}

\vspace{-2mm}
\subsubsection{\revise{Varying the Amount of Transitivity Violations.}}
\revise{
We perform a controlled experiment to evaluate how different methods behave when we vary the amount of transitivity violations in the ground-truth. From our survey on 30 real-world datasets, transitivity is violated when the ground-truth is incomplete or incorrect. Therefore, we can perform this experiment by corrupting the ground-truth labels to make the set of matching pairs incomplete or incorrect.
}

\revise{
The Camera and Monitor datasets have the highest number of labeled tuple pairs involved in transitivity constraints in the ground-truth, while other datasets have fewer such tuple pairs (e.g., the ground-truth matching pairs in the FZ dataset are all one-to-one mappings so no tuple pairs are involved in transitivity). In addition, the Camera and Monitor datasets are "perfectly" labeled in the sense that initially there are no transitivity violations in the ground-truth. Therefore, we use the Camera and Monitor datasets for this experiment. We inject transitivity violations with the following steps: We first randomly select a tuple $t_i$ that has matches in the ground-truth. Next, by a probability of $60\%$ we randomly mark one true matching pair that involves $t_i$ as a non-match, and in the remaining $40\%$ of cases we randomly choose a non-matching pair that involves $t_i$ and  mark it as a spurious match. These probabilities are selected based on our survey on real-world datasets in Section~\ref{sssec:survey_trans}. We repeat the two steps $xN_{\text{gt}}$ times, where $N_{\text{gt}}$ is the total number of matches in the original ground-truth; and $x$ controls the amount of violations we introduce (from 0 to 0.5 with a step size of 0.1). We compare with the two best performing baseline methods from Table~\ref{tbl:performance}: Snorkel and Majority Vote. We report the averaged score of all methods over the two datasets in Table~\ref{tbl:vary_trans}. As $x$ increases, the performance of all methods decreases because transitivity violations are achieved by corrupting the ground-truth labels. However, SIMPLE-EM always performs better than the two other baseline methods. 
}

\begin{table}[h]
\vspace{-3mm}
\color{revision}
\caption{\revise{F1-score of top methods when varying the amount of transitivity violations in the ground-truth (x).}}
\vspace{-3mm}
\resizebox{0.7\columnwidth}{!}{
\begin{tabular}{|l|l|l|l|l|l|l|}
\hline
$x$         & 0     & 0.1   & 0.2   & 0.3   & 0.4   & 0.5   \\ \hline
SIMPLE-EM & 0.880 & 0.841 & 0.802 & 0.764 & 0.726 & 0.697 \\ \hline
SN        & 0.815 & 0.777 & 0.740 & 0.705 & 0.670 & 0.636 \\ \hline
MV        & 0.786 & 0.750 & 0.715 & 0.681 & 0.648 & 0.616 \\ \hline
\end{tabular}}
\label{tbl:vary_trans}
\vspace{-3mm}
\end{table}

\subsection{Adaptation to Data Shift}
Data shift is a challenging problem in many real-world applications~\cite{quinonero2008dataset}. %
In this section, we study the behavior of LFs and traditional manual labeling during data shift.

In our datasets in Table~\ref{tbl:datasets}, we have two pairs of datasets with the same schema: (DA, DS) and (AB, AG). \revision{In addition, we have (AB, WA) where the attributes of WA is a super-set of the attributes in AB. We construct a new dataset WA' from WA using the subset attributes appearing also in AB so that the constructed dataset WA' has the same schema with AB.
}
To simulate data shift, we consider the following source-target data shift: DA-DS (shift from DA to DS),  AB-AG (shift from AB to AG), \revise{AB-WA' (shift from AB to WA')}.  
We use the following settings for LFs and manual labeling: %
\begin{itemize}[wide,labelindent=0pt]
\vspace{-1mm}
\item \textit{LFs.} When developing LFs in Section~\ref{ssec:lf_dev}, we reused many LFs from prior datasets. For each source-target dataset pair, we count the total number of LFs $N_1'$ in the target dataset. Since some of the LFs are reused from the source dataset, we count the number of newly developed LFs $N_2'$ for the target dataset. We report $\frac{N_1'-N_2'}{N_1'}$ as the amount of saved effort for labeling the target dataset. 
\item \textit{Manual labeling.} For each source-target dataset pair, we consider the task of learning on the target dataset. First, we ignore the source dataset and only consider the target dataset, and we use the active learning method. We count the number of labels $N_1$ queried by active learning when it reaches the performance of LFs (or its peak performance if it cannot match LFs). Second, we add all labeled data from the source dataset to the training set of active learning on the target dataset. Since the source and target dataset has the same schema, the features generated by Magellan~\cite{konda2018magellan} are the same, so we can directly use the labeled data from the source dataset to train a model for the target dataset. We then count the number of additional labels $N_2$ from the target dataset queried by active learning when it reaches the performance of LFs (or its peak performance if it cannot match LFs) on the target dataset. We report $\frac{N_1-N_2}{N_1}$ as the amount of saved effort for labeling the target dataset by using the manual labels in the source dataset. %
\end{itemize}

The results are shown in Table~\ref{tbl:data_shift}. In manual labeling, the labeled data on the source dataset is not always helpful for the target dataset. For example, on the AB-AG datasets \revise{and the AB-WA' datasets}, the saved labeling effort is negative. This means, when using labeled data from the source, one actually needs to label more data on the target dataset to achieve the same performance as in the case where one simply ignores the source dataset. This is understandable as the definition of being a match can be different on a different dataset. 
 Even when the labeled data from the source dataset is helpful (e.g., on the DA-DS datasets), the saved labeling effort is still significantly smaller than that for using LFs. In contrast, using LFs is more flexible and one can easily adapt to the new definition of being a match on the new dataset by reusing a subset of original LFs or by adding more LFs.  
 
In addition, we highlight that when the source and target datasets have different schemas (features are different), there is no way for manual labeling to reuse existing labels from the source dataset; However, for LFs, as long as there are some common or similar attributes, one can easily reuse the LFs written on these attributes.

\begin{table}[h]
\vspace{-3mm}
\caption{Saved labeling effort on the target dataset for LFs and manual labeling under data shift.}
\vspace{-4mm}
\renewcommand{\arraystretch}{0.8}
\resizebox{0.6\columnwidth}{!}{
\begin{tabular}{|c|c|c|}
\hline
     data shift& manual labeling & LFs \\ \hline
DA-DS&31.5\% &  62.5\% \\ \hline
AB-AG&-23.2\% &  63.6\% \\ \hline
\revise{AB-WA'}&  \revise{-9\%} & \revise{73.3\%} \\ \hline
\end{tabular}
}
\label{tbl:data_shift}
\vspace{-4mm}
\end{table}

\subsection{Sensitivity Analysis}
We analyze labeling model performance under varying sets of LFs.

\noindent\textbf{LF Randomization.} 
Some LFs use threshold values to assign labels (e.g., the name\_overlap LF in Figure~\ref{fig:example_lfs} compares the score variable to two different threshold values).
To assess LF sensitivity, we generate a new set of LFs from the original LFs by randomly tweaking the threshold values (if any) within a range around the original thresholds. 
 Additionally, we also take a random sample of the full set of LFs to see how methods perform with fewer LFs. 

\noindent\textbf{Results.} Table~\ref{tbl:sensitivity} shows the sensitivity of our method and baseline methods to changes in the LFs. Our method outperforms all baselines in all scenarios and generally shows less of a degradation in performance than other methods at 80\% and 60\% LF sampling. When the sampled proportion of LFs decreases, the performance improvement of our method over baseline methods decreases. This is because when there are fewer LFs, the benefit of carefully combining them decreases, as also reported in prior work~\cite{2018-VLDB-Snorkel-System.pdf}.

\begin{table}[h]
\vspace{-2mm}
\caption{Sensitivity to LFs. RT denotes randomized thresholds. $x\%$ denotes sampling $x\%$ of the original set of LFs. \revise{The original number of LFs for each dataset can be found in Table~\ref{tbl:lf_time_spent}. The scores are F1-scores averaged over all datasets. }}
\vspace{-3mm}
\renewcommand{\arraystretch}{0.8}
\color{revision}
\resizebox{0.85\columnwidth}{!}{\begin{tabular}{|l|l|l|l|l|l|}
\hline
     & Original & RT+100\% & RT+80\% & RT+60\% &RT+40\% \\ \hline
SIMPLE-EM   &  0.861     &0.856                    & 0.831                  &0.766  & 0.570          \\ \hline
MV   &  0.755     &0.718                    & 0.599                  &0.554  &0.537 \\ \hline
D\&S &0.717          &  0.672           & 0.595      &0.474       & 0.397 \\ \hline
EBCC   &  0.677     &0.632                    & 0.638                  &0.533  &0.504           \\ \hline
SN   &  0.772     &0.698                    & 0.674          &0.623  &0.554           \\ \hline
FS   &  0.553     &0.486                    & 0.460                  &0.447  & 0.445          \\ \hline
\end{tabular}}
\label{tbl:sensitivity}
\end{table}

\subsection{Truth Inference on General Tasks} 
We evaluate our proposed method (SIMPLE) from Section~\ref{sec:proposed_truth_inference} on general weak supervision tasks to verify that our proposed method works beyond EM. We consider all ten binary classification datasets in the WRENCH weak supervision benchmark repository~\cite{github_wrench, zhang2021wrench}. For all datasets, we use the provided LFs and performance metric from the benchmark~\cite{github_wrench, zhang2021wrench}. 
The results are shown in Table~\ref{tbl:truth_inference_on_wrench}. %

\begin{table}[h]
\vspace{-3mm}
\caption{Performance on the wrench benchmark~\cite{zhang2021wrench}. "F1" denotes F1-score and "acc" denotes accuracy score.}
\vspace{-3mm}
\setlength{\tabcolsep}{2.5pt}
\renewcommand{\arraystretch}{0.8}
\resizebox{\columnwidth}{!}{\begin{tabular}{|l|l|l|l|l|l|l|l|l|}
\hline
  Datasets      & \# of LFs & metric & SIMPLE & MV    &D\&S   & EBCC        & FS        & SN   \\ \hline
basketball      & 4 &F1           & 0.171  & \textbf{0.181}        & 0.171          & 0.171          &           0.171&  0.144        \\ \hline
commercial      &4 &F1              & 0.837   &0.846        &  0.778          & 0.775          & 0.763   &       \textbf{ 0.878 }     \\ \hline
tennis      &6 &F1              &  0.844  & \textbf{0.847}      & \textbf{0.847}          & \textbf{0.847}          & \textbf{0.847}         &            0.841 \\ \hline
yelp      & 8&acc       &  \textbf{0.744}  &0.722       & 0.683          &0.696           & 0.709   &     0.696        \\ \hline
imdb      &8 &acc         & \textbf{0.750}    & 0.737     &  0.744        &0.744          & 0.744    &        \textbf{0.750}          \\ \hline
spouse       &9 &F1    & \textbf{0.517}    & 0.492    &  0.343         & 0.343          &       0.505    & 0.455        \\ \hline
youtube     &10  & acc     & \textbf{0.916}    & 0.853     &   0.452        & 0.452            &   0.845  &       0.847     \\ \hline
cdr     &33  & F1      & \textbf{0.713}    & 0.672     &   0.001        & 0.087           &   0.104  &       0.666      \\ \hline
sms     &73  & F1      & 0.825    & 0.838     &   0.650        & 0          &   0  &       \textbf{0.840}       \\ \hline
census     &83  & F1      & \textbf{0.527}    & 0.330     &   0.001        & 0            &   0.209  &       0.445     \\ \hline
Avg. & -  &-      &  \textbf{0.684}    &0.652      &   0.467        & 0.412            &  0.490 &     0.656       \\ \hline
\end{tabular}}
\label{tbl:truth_inference_on_wrench}
\vspace{-1mm}
\end{table}

Our proposed method outperforms the best baseline on average by 3\%. This is significant considering that the best prior work only outperforms Majority Vote by 0.4\%.
It's quite surprising that our method obtains such good performance in spite of its simplicity.
In contrast, existing methods typically use very complicated models (e.g., graphical models~\cite{fu2020fast} and matrix completion models~\cite{ratner2019training}). %

\vspace{-2mm}
\section{Related Work}
\textbf{Entity Matching.}
\revise{
Supervised learning algorithms achieve the best results on entity matching~\cite{2016-VLDB-Magellan.pdf,2018-VLDB-DeepER.pdf,2018-SIGMOD-DeepMatcher-Design-Space-Exploration.pdf,2021-VLDB-Ditto,zhao2019auto}. However, they require large amounts of labeled data.
To reduce training data size in supervised EM, active learning based approaches selectively label useful tuple pairs, but need to have human annotators involved in the ML training process~\cite{2010-SIGMOD-AL-EM-MSR.pdf,2002-KDD-Sarawagi.pdf}.
Transfer learning is adopted to reuse information from existing datasets or pretrained language models~\cite{2021-VLDB-Ditto}, but is not robust when the target dataset is very different from source datasets or language models as we have shown in our experiments. 
Different from existing work, our work programmatically generates labels by adapting weak supervision (that has been successfully applied to label generation for general ML tasks~\cite{2016-NIPS-data-programming.pdf,2018-VLDB-Snorkel-System.pdf, snubaVLDB18}) to EM. Our work offers a new method to perform entity matching when no labeled data is available. 
}

\noindent\textbf{Truth Inference.}
\revise{The existing truth inference methods are typically designed for general tasks~\cite{2019-ICML-Exploiting-Worker-Correlation-for-Label-Aggregation.pdf,li2014resolving}. Our SIMPLE-EM method is specifically designed for EM by incorporating the transitivity property of EM and achieves better performance than general truth inference methods. In addition, existing methods are mostly hand-crafted, complicated models with various assumptions to implicitly restrict the hypothesis space to avoid trivial solutions~\cite{2019-ICML-Exploiting-Worker-Correlation-for-Label-Aggregation.pdf,li2014resolving, 2016-NIPS-data-programming.pdf,2018-VLDB-Snorkel-System.pdf}. In contrast, our method is based on a generic classifier and makes no assumptions. Our method restricts the hypothesis space in an explicit data-driven fashion through cross validation. The two distinctions of our method enabled our method to achieve better results on both general tasks and EM tasks. 
}

\vspace{-1mm}
\section{Conclusion}
In this work, we present a labeling model to generate high-quality EM labels by combining the predictions of different labeling functions in a weak supervision setting. 
We first propose a simple and powerful general labeling model for general weak supervision classification tasks. We then tailor the method to the task of EM by ensuring the predicted labels satisfy the transitivity property of EM.
Finally, we experimentally validate that our general labeling model works well on ten weak supervision datasets and find that the tailored version for EM is significantly better than existing approaches across several diverse datasets.

\bibliographystyle{ACM-Reference-Format}
\bibliography{Refs.bib}

\if\TR 1
\section{Appendix}
\label{sec:app}

\subsection{Duplicate-free Detection with Weak Supervision}
\label{app:dup_free}

We propose a method to detect whether each table is duplicate-free in two-table EM. 
The method uses information from the predicted matches (pairs of left tuple and right tuple) of a labeling model (without considering transitivity). Since the user has already written LFs for LR pairs, this method requires no additional effort from the user.

\noindent\textbf{Formulation of the Detection Method.}
Let $M=\{(t_{l_1}, t_{r_1}), \dots \}$ denote the set of found matches by the labeling model (without considering transitivity). Let $d_r$ denote the number of distinct right tuples in $M$. Consider the case when the predictions in $M$ are correct with a precision of 1. When the left table is duplicate-free we have $d_r=|M|$, and when the left table is not duplicate free we have $d_r<|M|$ as illustrated in Figure~\ref{fig:dup_free}. 
However, in practice, even when the left table is duplicate-free we may likely observe $d_r<|M|$ because the found matches $M$ can contain noise. Our idea is to test whether the observed data of $d_r<|M|$ can be explained by the noise in $M$, and if not, this indicates that the left table is not duplicate-free. (Note that if the observed data is $d_r=d_l=|M|$, tuple pairs in $M$ are not involved in transitivity constraints, so whether we enforce transitivity or not makes no difference.)

\begin{figure}[htb!]
  \centering
  \includegraphics[width=0.8\linewidth]{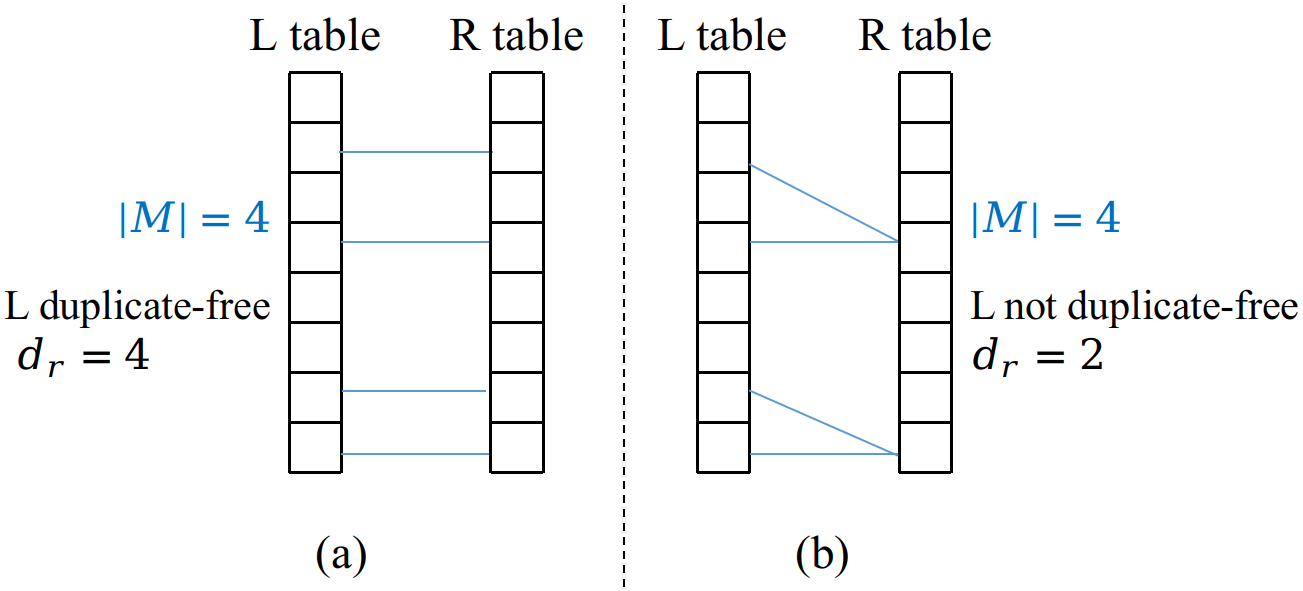}
  \caption{Illustration of matching edges when (a) L table is duplicate-free and (b) L table is not duplicate-free.}
  \label{fig:dup_free}
\end{figure}

Without loss of generality, we design a hypothesis test to detect whether the left table is duplicate-free.
Our observed data is $d_r$ and $M$, and we conduct the test when we observe that $d_r<|M|$.
Our null hypothesis is that the left table is duplicate-free. Let $x$ denote the number of true positives in $M$ and $|M|-x$ denote the number of false positives (Since the ground-truth is unknown, $x$ is a hidden variable.). Under the null hypothesis, the $x$ true positives are $x$ tuple pairs
with $x$ distinct right tuples, so $d_r<|M|$ is caused by the $|M|-x$ false positives. The labeling model makes mistakes (false positive predictions) when it encounters "unexpected" tuple pairs which usually do not follow the patterns that the set of LFs were designed for. Since these tuple pairs can be "unexpected" in random ways, these tuple pairs can be seen to be distributed randomly. Accordingly, the $|M|-x$ right tuples in the $|M|-x$ false positive pairs are distributed randomly and can be seen to be randomly selected from the right table (with replacement). By formulating this random process, we are able to obtain a distribution $p(d_r)$.
If the observed $\hat{d_r}$ is too small to be observed under the null hypothesis, we reject the null hypothesis. Formally, we reject the null hypothesis when:
\begin{equation}
\label{eq:dup_free_reject}
p(d_r<\hat{d_r})=\sum_{d_r=x}^{\hat{d_r}-1}p(d_r)<c
\end{equation}
where $c$ is a confidence level constant typically chosen as $0.05$.

\noindent\textbf{Computing Equation~\ref{eq:dup_free_reject}.}
There is a hidden variable $x$ in $p(d_r)$.
We would like to obtain the maximum likelihood estimation of $x$, which then can be used to compute Equation~\ref{eq:dup_free_reject}. Under our formulated random process, it can be shown that:
\begin{equation}
\label{eq:ndv_pdf}
p(d_r)=\sum_{i=d_r-x}^{|M|-x}{|M|-x\choose i}\frac{S_2(i,d_r-x)(N_r-x)!x^{|M|-x-i}}{N_r^{|M|-x}(N_r-d_r)!}
\end{equation}
where $S_2(n, k)$ is the number of ways to partition $n$ objects into $k$ non-empty sets and is known as the Sterling number of the second kind~\cite{Stirling}.
Since Equation~\ref{eq:ndv_pdf} is too complicated to be used in practice, we design our method to bypass the computation of Equation~\ref{eq:ndv_pdf}. We omit the derivation of Equation~\ref{eq:ndv_pdf} as it is not used in our method. 

Directly maximizing $p(d_r)$ with respect to $x$ in Equation~\ref{eq:ndv_pdf} is difficult. 
We first show in some cases we can make the decision of rejecting the null hypothesis without needing to find $x$. Intuitively, $d_r$ achieves the maximum value $d_r=|M|$ when $x$ is at its maximum $|M|$. When $x$ is smaller, more items are selected at random, and $d_r$ will be smaller. Therefore, we would expect:
\begin{equation}
\label{eq:upper_bound}
p(d_r<\hat{d_r}, x) \leq p(d_r<\hat{d_r}, x=0) = \sum_{d_r=0}^{\hat{d_r}-1} p(d_r, x=0)
\end{equation}
where $p(d_r, x=0)$ is the distribution of the number of distinct values $d_r$ in a set of size $|M|$ where each element is randomly selected from a set of size $N_r$ with replacement. It is given as~\cite{coverage_distri}:
\begin{equation}
\label{eq:p_dr_x_0}
p(d_r, x=0) = \frac{S_2(|M|,d_r)N_r!}{N_r^{|M|}(N_r-d_r)!}
\end{equation}
Intuitively, $\frac{S_2(|M|,d_r)N_r!}{(N_r-d_r)!}=\frac{S_2(|M|,d_r)d_r!N_r!}{(N_r-d_r)!d_r!}=S_2(|M|,d_r)d_r!{N_r \choose d_r}$ is the number of ways to select $|M|$ elements with exactly $d_r$ distinct values. $N_r^{|M|}$ is the total number of ways to select $|M|$ elements. Therefore, their division is the probability of having $d_r$ distinct values. 

Based on Equation~\ref{eq:upper_bound} and Equation~\ref{eq:p_dr_x_0}, we obtain an upper-bound of $p(d_r<\hat{d_r},x)$ as:
\begin{equation}
\label{eq:upper_bound}
p(d_r<\hat{d_r},x) \leq  \sum_{d_r=0}^{\hat{d_r}-1} \frac{S_2(|M|,d_r)N_r!}{N_r^{|M|}(N_r-d_r)!}
\end{equation} 
When the right hand side is smaller than $c$, for sure $p(d_r<\hat{d_r}, x)<c$, so we reject the null hypothesis. Otherwise, we have to find $x$ to make a decision. We resort to a simulation based approach. Specifically, we vary $x$ from 0 to $|M|$ with a step size of $|M|/10$ and, for each $x$, we perform the following simulation: we initialize a bag with $x$ unique numbers $\{1, 2 ,..., x\}$, then randomly select $|M|-x$ numbers in range $[1,N_r]$ to the bag with replacement (this is equivalent to generating $|M|-x$ random integers and can be done efficiently with numpy), and finally find the number of distinct numbers $d_r$ in the bag. By repeating this simulation many times, we obtain an empirical distribution $\hat{p}(d_r)$. We choose the value of $x$, so that the probability of the observed $\hat{d_r}$ is maximized under the empirical distribution, i.e. $\hat{p}(\hat{d_r})$ is maximized. We further perform the hypothesis test with the empirical distribution under the chosen $x$, i.e. we reject the null hypothesis when $\hat{p}(d_r<\hat{d_r})<c$. The time complexity of computing the upper-bound in Equation~\ref{eq:upper_bound} is $O(\hat{d_r})$ and the time complexity of the simulation is $O(|M|)$. 

\noindent\textbf{Discussion.}
With the hypothesis test, we are able to detect whether the left table is duplicate-free. By switching the left and right table and repeating the hypothesis test, we are equivalently able to detect whether the right table is duplicate-free. 
Note that obtaining the hidden variable $x$ by maximizing the likelihood of observed data biases the test toward the observed data and toward not rejecting the null hypothesis. This means we only reject the null hypothesis (left table is duplicate free) when the left table is significantly not duplicate free. This is intuitively acceptable, as when the left table only has a few duplicates, it is close to being duplicate-free and enforcing transitivity using the exact solution can still be helpful. We further empirically verify this in our experiments. 

\noindent\textbf{Experimental Evaluations of Duplicate-free Detection.}
We evaluate the effectiveness of our duplicate-free detection method on two-table datasets. The results are shown in Table~\ref{tbl:dup_free}. Each cell in the first two columns shows the number of duplicates in the L and R table in the ground truth and in the predicted set of matches $M$ of the labeling model without considering transitivity. Note the ground-truth and the predicted set of matches only include cross-table tuple pairs, so the number of duplicates is estimated based on the cross table matching pairs. For example, when $(t_{l_1}, t_{r_1})$ and $(t_{l_1}, t_{r_2})$ are two cross-table matching pairs, we know that $(t_{r_1}, t_{r_2})$ is a matching/duplicate pair in the right table. Since the three datasets IR, YY and ABN only includes a small fraction of the ground-truth of the cross-table matching pairs, the estimated number of duplicates is not accurate so we don't show them in Table~\ref{tbl:dup_free}.
The third row shows the duplicate-free detection results for each dataset using the method at the end of Section~\ref{ssec:trans_two_table}. The method is able to correctly detect that the Fodors-Zagats, DBLP-ACM, and Abt-Buy datasets are duplicate-free or almost duplicate-free. Note that the detection method uses the information in $M$ which actually contains many duplicates in the L table or R table in DBLP-ACM and Abt-Buy, but our proposed method is able to detect the two tables of the two datasets are duplicate-free. 
In addition, the AG dataset has fewer duplicates in $M$ than the DA dataset, which make it seem to be more likely to be duplicate-free than DA. However, our proposed method is able to judiciously recognize that AG is not duplicate free while DA is. 
The fourth row shows whether using the closed-form solution derived for the duplicate-free scenario is helpful for each dataset. We consider applying transitivity to be helpful when F1 score increases by applying transitivity (see Section \ref{ssec:ablation-study} for full ablation results). We can see whenever the detection method detects duplicate-free tables, using the closed-form solution derived for the duplicate-free scenario for that dataset is helpful, even though the dataset may not be completely duplicate-free (e.g., Abt-Buy). This also verifies our intuition at the end of Section~\ref{ssec:trans_two_table} that bias toward not rejecting the hypothesis of duplicate-free is acceptable.

\begin{table}[h]
\caption{Duplicate-free detection on two-table datasets.}
\setlength{\tabcolsep}{3.5pt}
\resizebox{\columnwidth}{!}{\begin{tabular}{|c|c|c|c|c|c|c|c|c|c|}
\hline
     & FZ & DA& DS & AB &AG & WA &IR &YY &ABN \\ \hline
\begin{tabular}{c}Ground truth \# dups \\ in (L, R)\end{tabular}    &  0, 0     &0, 0 & 2939, 129                   &16, 5 & 187, 9  &162, 8 & -&- &-       \\ \hline
\begin{tabular}{c}Predicted \# dups\\ from $M$ in (L, R)\end{tabular}    &  3, 3     &781, 817                 & 3973, 979                 &165, 172  &544, 355 &1265, 327 &- &- &- \\ \hline
\begin{tabular}{c}Dup-free prediction\\ in (L, R)\end{tabular}      &  T, T     &T, T                    & F, F                  &T, T  &F, F   &F,F &T,T &F,F &T,T        \\ \hline
\begin{tabular}{c}Is dup-free based\\solution helpful? \end{tabular}  &  Yes     &Yes                    & No                  &Yes  &No & No&Yes &No &Yes          \\ \hline
\end{tabular}}
\label{tbl:dup_free}
\end{table}

\fi
\end{document}